\documentclass[11pt]{article}

\input{epsf}
\usepackage{epsfig}
\usepackage{amssymb}
\usepackage{amsfonts}
\usepackage{amsbsy}
\usepackage[all]{xy}
\usepackage{amsmath}

\usepackage{amssymb,amscd}
\usepackage{mathrsfs}
\usepackage{amsmath,amsthm}
\usepackage{dsfont}
\usepackage{soul}
\usepackage{comment}

\def\be{ \begin{eqnarray} }
\def\ee{ \end{eqnarray}}

\def\Co0{{\rm Co}_0}

\def\exp{{\rm exp}}

\def\I{{\rm i}}

\def\log{{\rm log}}

\def\Tr{{\rm Tr}}

\def\p{\partial}

\def\one{{\hbox{ 1\kern-.8mm l}}}

\def\CA{{\cal A}}

\def\CD {{\cal D}}

\def\CF {{\cal F}}

\def\CH {{\cal H}}

\def\CJ {{\cal J}}

\def\CM {{\cal M}}
\def\CN {{\cal N}}
\def\CO {{\cal O}}

\def\CV {{\cal V}}

\def\CO {{\cal O}}
\def\CZ {{\cal Z}}

\def\CH {{\cal H}}

\def\CQ {{\cal Q}}
\def\CS {{\cal S}}

\def\CZ{{\cal Z}}

\def\IC{\mathbb{C}}

\def\IR{{\mathbb{R}}}

\def\IZ{{\mathbb{Z}}}

\def\fu{\mathfrak{u}}

\def\rmk#1{\bigskip\noindent{\bf Remarks} }

\usepackage{tikz}
\usetikzlibrary{arrows}
\usetikzlibrary{arrows.meta}
\usetikzlibrary{positioning}
\usetikzlibrary{shapes,snakes}
\usetikzlibrary{fit}

\def\lm{\limits}
\def\nn{\nonumber}

\def\lm{\limits}

\hoffset= - 0.9in    

\numberwithin{equation}{section}
\numberwithin{theorem}{section}

\title{BPS Hall Algebra of Scattering Hall States}
\author{Dmitry Galakhov\footnote{e-mails: galakhov@berkeley.edu; galakhov@itep.ru}}
\date{}

\begin{document}
\maketitle

\vspace{-1cm}
\begin{center}
	{\it Center for Theoretical Physics, University of California, Berkeley, USA, 94720}\\
	{\it Institute for Information Transmission Problems, Moscow, Russia, 127994}
\end{center}

\begin{abstract}
	Starting with a very pedestrian point of view we compare two different at the first glance definitions for an algebra associated to BPS states in supersymmetric fields theories. One proposed by Harvey and Moore exploits $S$-matrices of BPS states as structure constants of a new algebra. Another one proposed by  Kontsevich and Soibelman gives a  construction according to the structure of cohomological Hall algebras. We show these two constructions give equivalent algebras.
\end{abstract}

\tableofcontents

\section{Introduction and discussion}

A purpose of this note is to compare two BPS algebras derived as a scattering algebra of states and as a cohomological Hall algebra.

A special class of states saturating Bogomolny-Prasad-Sommerfeld (BPS) bound has  attracted an interest of both physicists and mathematicians over recent years. For a review see \cite{Cecotti:2012se, Denef:2007vg, Kontsevich:2008fj, Pioline:2011gf,Reineke1,Brennan:2018ura,Moore:2015szp}. They represent an affordable example of a family of quantum states in theories with supersymmetries that admits simultaneously an exact non-perturbative calculation of some corresponding quantum numbers and non-trivial dynamics.

String theory proposes an efficient and rather universal description of BPS states in terms of D-branes (see, for example, \cite{Strominger:1996sh}), whose IR dynamics \cite{Douglas:1996sw} gives rise to gauge theories admitting quiver description. There are multiple examples (see, for instance, \cite{Denef,Aganagic:2010qr,Cirafici:2008sn, Alim:2011kw} and references therein) in the literature implementing a quiver description to derive right counting of BPS states of the corresponding initial gauge or gravity theory.

Here we will just concentrate on a rather rich class of so called class $\CS$ models \cite{Witten:1997sc,Gaiotto:2009we,Gaiotto:2009hg}. Class $\CS$ theories are known to have a nice low energy effective behavior \cite{Seiberg:1994rs} when Higgs fields acquire vevs and the gauge group is broken to a factor  $U(1)^r$. In this description BPS states are represented by dyonic particles charged both electrically and magnetically. So that the charges are taking values in an integral charge lattice $\IZ^{2r}$. An effective model for such particles can be given in terms of  quiver $\CN=4$ supersymmetric quantum mechanics (SQM)\cite{Denef}. Quiver SQM is just a simple dimensional reduction of $\CN=1$ 4d super-Yang-Mills theory with a matter-field content encoded in a quiver. The choice of the corresponding quiver is dictated by a charge lattice and Dirac-Shwinger-Zwanziger (DSZ) pairing on it, moduli dependence of the theory descends to Fayet-Illiopolous (FI) stability parameters.

A remarkable property of BPS spectra in many models is a wall-crossing phenomenon: a theory may have parameters -- moduli, and BPS spectra are only piecewise constant as functions of the moduli breaking the moduli space into stability chambers all having different BPS spectra. To count BPS states one usually implements characters of Hilbert spaces of BPS states. A quite useful quantity for 4d $\CN=2$ gauge and gravity theories is so called protected spin character (PSC) \cite{Gaiotto:2010be}, or a refined BPS index \cite{MPS}, defined as
$$
\Omega(y)=\Tr_{H_{\rm BPS}}y^{2J_3}(-y)^{2I_3}
$$
where $J_3$ and $I_3$ are $SU(2)_J$ spin and $SU(2)_R$ isospin Cartan generators correspondingly, and the trace is taken over the BPS Hilbert space. PSCs are functions of the moduli of the theory and they are piecwise constant due to wall-crossing phenomena.

In the literature \cite{Gabella:2017hpz} there is a construction of a map from the class $\CS$ theory with moduli near a marginal stability wall to the corresponding effective quiver description of BPS states in this theory. As well one may use localization techniques \cite{Ohta:2014ria,Ohta:2015fpe,Cordova:2014oxa} to calculate corresponding PSCs in this effective quiver SQM.

Supersymmetric theories are, in general, a nice source of various algebraic constructions. A natural choice here is to consider a subalgebra of operators mapping BPS states to BPS states. Since the BPS Hilbert spaces are naturally graded by charge lattice vectors and the graded subspaces are usually finite dimensional there is a good hope that the corresponding algebra turns out to be finitely or at least countably generated. There are various examples of algebras constructed in this way even including non-local operators, see for example \cite{Nekrasov:2012xe,Kimura:2016dys,Bullimore:2016hdc,Awata:2017cnz}. Non-local operators are allowed to change numbers of BPS particles and perform maps between Hilbert spaces with different gradings. Eventually, one may consider the BPS Hilbert space as an algebra itself with a multiplication:
$$
m:\quad H_{\rm BPS}\otimes H_{\rm BPS} \to H_{\rm BPS}
$$

We will compare two approaches to constructing such an algebra for $\CN=4$ quiver SQM: a physical one proposed by Harvey and Moore in \cite{Moore} and a mathematical one proposed by Kontsevich and Soibelman in \cite{KS}. The physical approach dictates one to consider two-to-one scattering processes of BPS states and use corresponding $S$-matrices as structure constants of the new algebra. In the mathematical approach authors relate the Hilbert space of BPS states to equivariant cohomologies of quiver representations then construct a cohomological Hall\footnote{Refers to the mathematician Philip Hall (1904 -- 1982).} algebra (CoHA) \cite{KS} (see \cite{Rapcak:2018nsl} for the most recent systematic review of the subject) as a natural map beteen corresponding cohomological spaces.

We will use approach of \cite{Witten} to localize the theory to the Coulomb branch following \cite{MPS,MPS2,Manschot:2011xc,Cordova:2014oxa}, further we localize it to the fixed points of the spin $J_3$-generators. Localized states are represented by bound molecules of dyons, where elementary dyons are confined on the third axis in $\IR^3$. The wave function will split into an spin-invariant part and an anti-holomorphic part. The anti-holomorphic part is represented by a modified version of Hall\footnote{Refers to the physicist Edwin Hall (1855 -- 1938).} wave functions \cite{Laughlin}: multi-particle states occupying Landau levels in an external magnetic field with certain permutation symmetry. 

We will construct Hilbert spaces of framed BPS states and consider their scattereing. We will show that the corresponding scattering algebra of Hall states coincides with the cohmological Hall algebra. Wall crossing formul{\ae} \cite{Kontsevich:2008fj,Gaiotto:2010be} describing transitions of BPS spectra across the walls of marginal stability and combinatorial formulas for PSCs \cite{Manschot:2011xc}  follow automatically.

As we have mentioned a quiver supersymmetric quantum mechanics description of BPS states is originating from certain Calabi-Yau compactifications
of type II string theory \cite{Denef} where BPS states take form of wrapped D-branes. A similar description has a quite wide range of applications including class $\CS$ theories, 5d theories and others \cite{Rapcak:2018nsl}. It would be natural to expect an existence of a similar descriptions for corresponding BPS algebras and, ideally, an explicit physically motivated construction for BPS algebras corresponding to generic shuffle algebras \cite{Feigin}. 

Hall algebras have a natural structure of bi-algebras \cite{Shifman1}. Considering an algebra as a module over itself one could identify multiplication and co-multiplication by a reference state of certain charge as raising/lowering operators of a chiral algebra \cite{Nekrasov:2012xe,Kimura:2016dys,Bullimore:2016hdc,Awata:2017cnz}. Relations inside this algebra lead to relations on a partition function counting BPS states maybe with some weights. Since usually these relations take a form of differential, more generally difference, equations, this leads to a spectral problem and spectral curves. Eventually, one might expect to construct a map inverse to \cite{Gabella:2017hpz}: to a given quiver one would be able to associate some ``minimal" spectral cover description of the BPS states. 

Despite we manage to show two definitions of BPS algebra in chosen series of models are equivalent we do not discuss a conceptual sourse of this equivalence. The construction implementing scattering of two BPS states proposes as a result in the product a new entangled state of two initial asymptotic states (for a recent review of entanglement in theories with an enhanced set of symmetries like TQFTs see \cite{Melnikov:2018zfn}). On the other hand the structure of the Hall algebra requires an implementation of extensions \cite{Shifman1,Shifman2}\footnote{I thank Tudor Dimofte for this valuable remark. }. A natural second step in this reasoning would be to try to clarify physically this relation between entanglement and extensions at least in the used family of models.

We leave  a deeper analysis and conceptual understanding of these problems to future work. 

This note is composed as follows. In Section \ref{QQM} we review some basics of quiver supersymmetric quantum mechanics proposed in \cite{Denef} as a machinery for an effective description of BPS states. In Section \ref{Hilbert} we do all the preliminary work to construct localized wave functions of the ground states. The heart of this note is an explicit wave function form \eqref{holowf} for the Hall state. Eventually, we derive anti-holomorphic wave functions and reproduce the Manschot-Pioline-Sen formula for the refined index \cite{MPS2}. Finally, in Section \ref{CoHA} we consider scattering of framed BPS states and derive the corresponding scattering algebra. In Appendix \ref{examples} some examples demonstrating the use of discussed techniques are collected.

\section*{Acknowledgments}
I am indebted to Gregory W. Moore for attracting my initial attention to problems in question.
I would like to thank Mina Aganagic, Semeon Arthamonov, Tudor Dimofte, Gregory W. Moore, Boris Pioline, Andrey Smirnov, Yan Soibelman, Masahito Yamazaki for fruitful discussions, valuable suggestions and remarks on different stages of this project.  I would like to thank The University
of California at Davis and The University of North Carolina at Chapel Hill for great hospitality during completion of this work. My research is supported by the Berkeley Center for Theoretical Physics, by NSF grant 1521446 and by the Simons Foundation through the Simons Investigatiors program. My work is supported in part by RFBR grants 16-01-00291, 18-31-20046-mol-a-ved and 19-51-53014 GFEN-a.

\section{Quiver quantum mechanics}\label{QQM}

In this section we will review briefly a setting of $\CN=4$ quiver supersymmetric quantum mechanics. For an indepth description the reader is referred to \cite{Denef,Ohta:2014ria}. One associates a quiver quantum mechanics Lagrangian to quiver data. Quiver data $(\mathscr{Q},\vec n,\vec \theta)$ contain three ingredients. The first ingredient $\mathscr{Q}$ is a quiver -- an oriented graph. We will denote the set of vertices of a given quiver as $\CV$ and the set of arrows as $\CA$. The second ingredient $\vec n=\{n_v\}_{v\in \CV}$ is a dimension vector of non-negative integers $n_v$ defining dimensions of complex vector spaces associated to quiver vertices. Finally, $\vec \theta=\{\theta_v \}_{v\in\CV}$ is a vector of real valued stability parameters. A quiver representation is defined as a quotient:
$$
\bigotimes\lm_{(a:v\to w)\in\CA}{\rm Hom}(\IC^{n_v},\IC^{n_w})\Big/
\prod\lm_{v\in \CV} GL(n_v,\IC)
$$

The Lagrangian of $\CN=4$ supersymmetric quiver quantum mechanics is constructed as a dimensional reduction of $\CN=1$ 4d super Yang-Mills theory where the field-matter content is encoded by the quiver data $(\mathscr{Q},\vec \gamma,\vec \theta)$. A vertex $v\in\CV$ defines vector multiplet $(A_v,X^i_v,\lambda_{\alpha,v},D_v)$ corresponding to the gauge action of $U(n_v)$. An arrow $a:v\to w$ defines a chiral multiplet $(q_a,\psi_{\alpha,a},F_a)$ bi-fundamental in $U(n_v)\times \overline{U(n_w)}$. Stability parameters $\theta_v$ define FI terms associated with each group factor $U(n_v)$. So the Lagrangian is just a sum of elementary Lagrangians associated to vertices and arrows:
\be\label{Lag}
L=\sum\lm_{v\in\CV}L_v+\sum\lm_{a\in\CA}L_a
\ee
The gauge symmetry of this Lagrangian is described by a group:
\be
G=\prod\lm_{v\in\CV}U(n_v)
\ee

The vertex Lagrangian reads \cite{Ohta:2014ria}:\footnote{Here and in what follows we use conventions for fermion notations from \cite{WessBagger}.}
\be
L_v=\Tr\left[\frac{1}{2}(\CD_0X_v^i)^2+\frac{1}{4}\left[X_v^i, X_v^j \right]^2-\I \bar\lambda_v\bar\sigma^0\CD_0 \lambda_v+\bar\lambda_v\bar\sigma^i\left[X_v^i,\lambda_v \right] +\frac{1}{2}D_v^2-\theta_v D_v  \right]
\ee

The covariant derivative in this case is defined as
\be
\CD_0 X_v=\p_0X_v+\I\left[A_v,X_v \right]
\ee

This Lagrangian is invariant up to total derivatives under the following SUSY transformations:
\be\label{SUSY}
\begin{split}
	\delta A_v=-\I \xi\sigma^0 \bar\lambda_v+\I  \lambda_v\sigma^0\bar\xi\\
	\delta X_v^i=\I \xi\sigma^i \bar\lambda_v-\I  \lambda_v\sigma^i\bar\xi\\
	\delta \lambda_v=\I\xi D_v+2\sigma^{0i}\xi \CD_0 X_v^i+\I \sigma^{ij} \xi\left[X_v^i,X_v^j \right]\\
	\delta D_v=-\xi\sigma^0\CD_0\bar\lambda_v-\I\xi\sigma^i\left[X_v^i,\bar\lambda_v \right] -\CD_0\lambda_v\sigma^0\bar\xi-\I \left[X_v^i,\lambda_v \right] \sigma^i \bar\xi
\end{split}
\ee

The arrow Lagrangian reads:
\be
\begin{split}
	L_{(a:v\to w)}=\Tr\left[ |\CD_0 q_a|^2-|X_w^iq_a-q_aX_v^i|^2 - \I \bar\psi_a\bar\sigma^0 \CD_0 \psi_a + \bar\psi_a \bar\sigma^i \left(X_w^i\psi_a-\psi_a X_v^i \right)+|F_a|^2+\right.\\
	\left. +\I\sqrt{2}\left((\bar q_a\lambda_w-\lambda_v\bar q_a)\psi_a-\bar\psi_a(\bar\lambda_w q_a-q_a\bar\lambda_v)\right)+\bar q_a(D_w q_a-q_a D_v) \right]
\end{split}
\ee
The covariant derivative in this case correspodns to the following expression:
\be
\CD_0 q_a=\p_0 q_a+\I (A_w q_a-q_a A_v)
\ee
And SUSY transformations of the fields are:
\be
\begin{split}
	\delta q_a=\sqrt{2}\xi\psi_a\\
	\delta\psi_a=\I\sqrt{2}\left( \sigma^0 \bar\xi \CD_0 q_a+\I \sigma^i \bar\xi (X_w^i q_a-q_a X_v^i) \right)+\sqrt{2}\xi F_a\\
	\delta F_a=\I\sqrt{2}\left( \bar\xi\bar\sigma^0 \CD_0 \psi_a+\I \bar\xi \bar\sigma^i (X_w^i\psi_a-\psi_aX_v^i)  \right)+2\I\bar\xi(\bar\lambda_w q_a-q_a\bar\lambda_v)
\end{split}
\ee

We will follow standard procedures allowing one to define the corresponding Hamiltonian and a Hilbert space of ground states. At the first step we should define momenta:
\be
\Pi_i=\CD_0X_i^{\dag},\quad \pi_a=\CD_0 \bar q_a,\quad \pi_a=\CD_0 q_a
\ee
Operators satisfy canonical commutation relations:
\be
\begin{split}
	\left[(X_i)^a_b,(\Pi_j)^c_d \right]=\I\delta_{ij}\delta^a_d\delta^c_b,\quad
	\left[q^c_d,\pi^e_f \right]=\I\delta_{ab}\delta^c_f\delta_d^e, \quad	\left[\bar q^c_d,\bar\pi^e_f \right]=\I\delta_{ab}\delta^c_f\delta_d^e\\
	\left\{(\bar\lambda_{\dot \alpha})^a_b,(\lambda_{\alpha})^c_d \right\}=-\sigma^0_{\alpha\dot\alpha}\delta^a_d\delta^c_b,\quad \left\{(\bar\psi_{\dot \alpha})^a_b,(\psi_{\alpha})^c_d \right\}=-\sigma^0_{\alpha\dot\alpha}\delta^a_d\delta^c_b
\end{split}
\ee

Having those we define supercharges as generators of the SUSY transforms according to the canonical rule:
$$
\delta \CO=\I\left[\xi Q+\bar\xi\bar Q,\CO \right]
$$

Supercharges read:
\be
\begin{split}
	Q_{\alpha}^{(v)}=\Tr\left[-\I(\sigma^0 \bar\lambda_v)_{\alpha} D_v-(\sigma^i\bar\lambda_v)_{\alpha}\left(\Pi_i+\frac{1}{2}\epsilon_{ijk}\left[X_v^j,X_v^k \right] \right) \right]\\
	\bar Q_{\dot\alpha}^{(v)}=\Tr\left[\I(\lambda \sigma^0)_{\dot\alpha} D_v-(\lambda_v\sigma^i)_{\dot\alpha}\left(\Pi_i-\frac{1}{2}\epsilon_{ijk}\left[X_v^j,X_v^k \right] \right) \right]\\
	Q_{\alpha}^{(a:v\to w)}=\I\sqrt{2}\Tr\left[ 
	\psi_{a\alpha}\pi_a-2\I(\bar q_a X^i_w-X^i_v\bar q_a) (\sigma^{i 0}\psi)_{\alpha}
	\right] \\
	\bar Q_{\dot\alpha}^{(a:v\to w)}=-\I\sqrt{2}\Tr\left[
	\bar\psi_{a\dot\alpha}\bar \pi_a+2\I(\bar\psi_a\bar\sigma^{0i})_{\dot\alpha}(X_w^iq_a-q_a X_v^i)
	\right]\\
Q_\alpha=\sum\lm_{v\in \CV}Q_{\alpha}^{(v)}+\sum\lm_{a\in \CA}Q_{\alpha}^{(a)},\quad
\bar Q_{\dot \alpha}=\sum\lm_{v\in \CV}\bar Q_{\dot\alpha}^{(v)}+\sum\lm_{a\in \CA}\bar Q_{\dot\alpha}^{(a)}
\end{split}
\ee

Fields $A_v$, $D_v$ and $F_a$  are non-dynamical, they do not have associated momentum operators. Instead they produce additional conditions imposed on the Hilbert space. Fields $A_v$ produce a gauge invariance condition for the wave functions, while $D_v$ acquire expectation values:
\be
D_v=\theta_v+\sum\lm_{a:v\to \bullet} \bar q_a q_a-\sum\lm_{a:\bullet\to v}q_a\bar q_a
\ee
In this note we restrict ourselves to quivers without superpotential, so $F_a=0$.

These supecharges form the following $\CN=4$ 1d superalgebra:
\be
\begin{split}
	\left\{ Q_{\alpha}, \bar Q_{\dot\alpha}  \right\} = -2\sigma^0_{\alpha\dot\alpha}\CH-2\I \sigma_{\alpha\dot\alpha}^{\mu} \sum\lm_{v\in\CV}{\bf g}_v(X_{v, \mu})\\
	\left\{Q_\alpha,Q_\beta \right\}=\left\{\bar Q_{\dot\alpha},\bar Q_{\dot\beta} \right\}=0
\end{split}
\ee
where $\CH$ is the Hamiltonian and ${\bf g}_v(\varphi)$ is a gauge transform generator corresponding to an algebra element $\varphi\in \fu(n_v)$, and we have used a 4d vector $X_{v,\mu}=(A_v,X_v^1,X_v^2,X_v^3)$.

In addition to supersymmetries the quiver quantum mechanics action is invariant under spin $SU(2)_J$ and phase $U(1)_R$ rotations. The corresponding charges are presented in the following table \cite{Ohta:2014ria}:
\be
\begin{array}{c|c|c|c|c|c|c|c}
	& A & X_i & \lambda_{\alpha} & D & q & \psi_{\alpha} & F \\
\hline
SU(2)_J & {\bf 1} & {\bf 3} & {\bf 2} & {\bf 1} & {\bf 1} & {\bf 2} & {\bf 1}\\
\hline
U(1)_R & 0 & 0 & 1/2 & 0& r & r-1/2 & r-1\\
\end{array}
\ee

Supercharges commute with the angular momentum operator in the following way:
\be
\left[J_k, Q_{\alpha} \right]=(\sigma^{k0}){}_{\alpha}{}^{\beta}Q_\beta,\quad 
\left[J_k, \bar Q_{\dot\alpha} \right]=\bar Q_{\dot\beta}(\bar\sigma^{k0}){}^{\dot\beta}{}_{\dot\alpha}
\ee
in particular:
$$
\left[J_3,Q_1 \right]=-\frac{1}{2}Q_1,\quad \left[J_3,Q_2 \right]=\frac{1}{2}Q_2,\quad \left[J_3,\bar Q_{\dot 1} \right]=\frac{1}{2}\bar Q_{\dot 1},\quad \left[J_3,\bar Q_{\dot 2} \right]=-\frac{1}{2}\bar Q_{\dot 2}
$$
and with $R$-charge in the following way:
\be
\left[R, Q_{\alpha} \right]=\frac{1}{2} Q_{\alpha},\quad \left[R, \bar Q_{\dot\alpha} \right]=-\frac{1}{2} \bar Q_{\dot\alpha}
\ee

Obviously, the angular momentum operator commutes with the Hamiltonian. So we may define a protected spin character as a weighted Witten index:
\be
\Omega(y):=\Tr(-1)^F y^{2J_3}
\ee
Here $H_0$ is a null-space of the Hamiltonian.

To calculate the protected spin character we will tend to a description of \cite{MPS}, where composed BPS states are represeted by ``molecules" of interacting elementary dyons. Further configuration spaces of these molecules are localized to $J_3$-invariant fixed points. We will clarify this prescription in what follows. Now let us say that a more rigorous approach of \cite{Ohta:2014ria} suggests that one should choose either of Hamiltonian representations: $\CH=\frac{1}{2}\{Q_1,\bar Q_{\dot 1} \}$ or $\CH=\frac{1}{2}\{Q_2,\bar Q_{\dot 2} \}$ and calculate either of manifest indices:
\be\label{PSC}
\Omega_+=\Tr (-1)^Fe^{\beta \frac{1}{2}\{Q_1,\bar Q_{\dot 1} \} } y^{J_3+ R},\quad 
\Omega_-=\Tr (-1)^Fe^{\beta \frac{1}{2}\{Q_2,\bar Q_{\dot 2} \} } y^{J_3- R}
\ee
Fortunately, due to the no exotics theorem \cite{Chuang:2013wt} the resulting BPS states are $R$-singlets. So without loss of generality we could choose either of supercharges to define a differential on the field space. Eventually, we will choose $Q_1$.

\section{Hilbert space of ground states in quiver quantum mechanics}\label{Hilbert}
\subsection{Preliminaries}\label{s31}
There are various approaches to use a localization in a theory with supersymmetry (see \cite{Pestun:2016zxk} for the most recent reviews of localization techniques). We will follow an authentic approach of \cite{Witten} allowing one to calculate an approximate Hilbert space of ground states. The particle is moving in a manifold $X$ endowed with an action of Lie group $G$. In this setup a supercharge has a form of a conjugated equivariant Cartan differential:
\be
\CQ=e^{-\sum\lm_i h_i}(d+\iota_v) e^{\sum\lm_i h_i}
\ee
where $h_i$ are $G$-invariant height functions. Then one considers a family of supercharges with rescaled height function and the Killing vector field:
\be\label{difff}
\CQ_{\vec s}=e^{-\sum\lm_i s_i h_i}(d+\iota_{s_0 v}) e^{\sum\lm_i s_i h_i}
\ee
Restricted to a $G$-invariant Hilbert space the Hamiltonian is just a Laplacian:
\be
\CH_{\vec s}=\frac{1}{2}\left(\CQ_{\vec s} \CQ_{\vec s}^\dag+\CQ_{\vec s}^\dag \CQ_{\vec s} \right)
\ee
following \cite{Witten} one expects the ground states to be in one-to-one correspondence with the $G$-equivariant cohmologies of $X$, therefore the ground Hilbert space is independent of  $\vec s$. 
The corresponding potential term in the Hamiltonian behaves as
$$
\sum\lm_i s_i^2 (\vec\nabla h_i)^2+s_0^2 \vec v^2 
$$

So the localization paradigm proposes to scale $\vec s$ to infinity where the wave functions are approximated to the first order by oscillatory modes around $G$-fixed critical points of $h_i$.

To use this technique in our case let us change coordinates:
\be
Z=X_1+\I X_2,\quad \bar Z=X_1-\I X_2,\quad \Pi_z=\frac{1}{2}(\Pi_1-\I \Pi_2),\quad \Pi_{\bar z}=\frac{1}{2}(\Pi_1+\I \Pi_2)
\ee

Then supercharges can be rewritten in a much simpler form:
\be\label{supercharges}
\begin{split}
	Q_1=e^{h_1+h_2}\left(\sum\lm_{v\in\CV}\Tr\left( -\bar\lambda_v^{\dot 1}\Pi_v^3-2\bar\lambda_v^{\dot 2}\Pi_v^z \right)+\I\sqrt{2}\sum\lm_{(a:v\to w)\in\CA}\Tr \left(   \psi_{a1}\pi_a+2\I(\bar q_a \bar Z_w-\bar Z_v\bar q_a)\psi_{a2}
	\right)  \right) e^{-h_1-h_2}\\
	Q_2=e^{-h_1+h_2}\left(\sum\lm_{v\in\CV}\Tr\left(\bar\lambda_v^{\dot 2}\Pi_v^3-2\bar\lambda_v^{\dot 1}\Pi_v^{\bar z} \right)+\I\sqrt{2}\sum\lm_{(a:v\to w)\in\CA}\Tr \left( \psi_{a2}\pi_a+2\I\sqrt{2}(\bar q_a Z_w- Z_v\bar q_a)\psi_{a1}
	\right)  \right) e^{h_1-h_2}\\
	\bar Q_{\dot 1}=e^{-h_1-h_2}\left(\sum\lm_{v\in\CV}\Tr\left( -\lambda_v^{1}\Pi_v^3-2\lambda_v^{2}\Pi_v^{\bar z}\right)-\I\sqrt{2}\sum\lm_{(a:v\to w)\in\CA} \Tr\left(   \bar\psi_{ a \dot1}\bar \pi_a-2\I\bar\psi_{ a\dot2}( Z_w q_a-q_a Z_v)
	\right)  \right) e^{h_1+h_2}\\
	\bar Q_{\dot 2}=e^{h_1-h_2}\left(\sum\lm_{v\in\CV}\Tr\left(\lambda_v^{ 2}\Pi_v^3-2\lambda_v^{1}\Pi_v^{z} \right)-\I\sqrt{2}\sum\lm_{(a:v\to w)\in\CA} \Tr\left(   \bar\psi_{ a \dot2}\bar\pi_a-2\I\bar\psi_{a \dot 1}(\bar Z_w q_a- q_a \bar Z_v)
	\right)  \right) e^{-h_1+h_2}\\
\end{split}
\ee
where 
\be
\begin{split}
	h_1=\sum\lm_{v\in\CV}\Tr X_v^3\left(\theta_v+\sum\lm_{a:v\to\bullet}\bar q_aq_a-\sum\lm_{a:\bullet\to v}q_a \bar q_a\right)\\
	h_2=\frac{1}{2}\sum\lm_{v\in\CV}\Tr X_v^3\left[Z_v,\bar Z_v \right]
\end{split}
\ee

Rewritten in this way the supercharges have explicitly the form of differential \eqref{difff}.
Without loss of generality we could choose $Q_1$ and $\bar Q_{\dot 1}$ as $\CQ$ and $\CQ^{\dag}$:
\be
\begin{split}
	\CQ=e^{{\bf s}_1 h_1+{\bf s}_2 h_2}\left(\sum\lm_{v\in\CV}\Tr\left( -\bar\lambda_v^{\dot 1}\Pi_v^3-2\bar\lambda_v^{\dot 2}\Pi_v^z \right)+\right.\\ \left.+\I\sqrt{2}\sum\lm_{(a:v\to w)\in\CA}\Tr \left(   \psi_{a1}\pi_a+2\I {\bf s}_3(\bar q_a \bar Z_w-\bar Z_v\bar q_a)\psi_{a2}
	\right)  \right) e^{-{\bf s}_1 h_1-{\bf s}_2 h_2}\\
	\CQ^\dag=e^{-{\bf s}_1h_1-{\bf s}_2h_2}\left(\sum\lm_{v\in\CV}\Tr\left( -\lambda_v^{1}\Pi_v^3-2\lambda_v^{2}\Pi_v^{\bar z}\right)-\right.\\ \left.-\I\sqrt{2}\sum\lm_{(a:v\to w)\in\CA} \Tr\left(   \bar\psi_{\dot a1}\bar \pi_a-2\I {\bf s}_3\bar\psi_{\dot a2}( Z_w q_a-q_a Z_v)
	\right)  \right) e^{{\bf s}_1h_1+{\bf s}_2h_2}\\
\end{split}
\ee

Then the corresponding family of Laplacians reads:
\be
\begin{split}
	\Delta_{\bf s}=\frac{1}{2}\left\{\CQ,\CQ^\dag \right\}=\\
	=\sum\lm_{v\in\CV} \Tr\left\{ \frac{1}{2}(\Pi_i^v)^2-\frac{{\bf s}_2^2}{4}\left[X^i_v,X^j_v \right]^2 -{\bf s}_2\bar\lambda_v \bar\sigma^i\left[X_v^i,\lambda_v \right] +\frac{{\bf s}_1^2}{2} \left( \theta_v+\sum\lm_{a:v\to \bullet} \bar q_a q_a-\sum\lm_{a:\bullet\to v}q_a\bar q_a \right)^2 \right\}+\\
	+\sum\lm_{(a:v\to w)\in\CA}\Tr\Big\{
	\bar \pi_a \pi_a+{\bf s}_1^2|X_w^3q_a-q_a X_v^3|^2+{\bf s}_3^2|Z_w q_a-q_a Z_v|^2-\\
	 -\bar\psi_a({\bf s}_3 \bar\sigma_1, {\bf s}_3 \bar\sigma_2, {\bf s}_1 \bar\sigma_3 ) \cdot\left(\vec X_w \psi_a-\psi_a \vec X_v \right)
	-\I\sqrt{2}{\bf s}_1\left( (\bar q_a\lambda_w^1-\lambda_v^1\bar q_a)\psi_{a1}-\bar\psi_{a\dot 1}(\bar\lambda_w^{\dot 1} q_a-q_a\bar\lambda_v^{\dot 1})
	\right)-\\ 
	-\I\sqrt{2}{\bf s}_3\left( (\bar q_a\lambda_w^2-\lambda_v^2\bar q_a)\psi_{a2}-\bar\psi_{a\dot 2}(\bar\lambda_w^{\dot 2} q_a-q_a\bar\lambda_v^{\dot 2})\right)
	\Big\}
\end{split}
\ee

Eventually one can introduce three types of moment maps labeling various critical points of $h_i$:
\be\label{mm}
\begin{split}
	\mu_{v,ij}^{(1)}=\left[X_v^i,X_v^j \right],\\
	\mu_v^{(2)}=\theta_v+\sum\lm_{a:v\to \bullet} \bar q_a q_a-\sum\lm_{a:\bullet\to v}q_a\bar q_a\\
	\mu^{(3)}_{a,i}=X_w^i q_a-q_a X_v^i\\
\end{split}
\ee
Then the set of $G$-fixed critical points is given by the standard quotient construction:
\be\label{locus}
\CM=\left(\bigcap\lm_I \mu^{-1}_I(0)\right)/G
\ee

It is easy to find corresponding $U(1)_{J_3+R}$-charges of corresponding fields just from a requirement that supercharge $Q_1$ \eqref{supercharges} remains invariant:
\be
\begin{array}{c|c|c|c|c|c}
	\bar z & z & \bar\lambda_{\dot 1} & \lambda_1 & \psi_2 & \bar \psi_{\dot 2}\\
	\hline
	-1 & 1 & 1 & -1& 1 & -1\\
\end{array}
\ee
Other charges are zero. 

\subsection{Localization on the Coulomb branch}
The localization on the Coulomb branch supports a multi-centered molecular picture we are aiming for, and it is defined by a particular regime of scaling parameters $\vec {\bf s}$. We assume ${\bf s}_3={\bf s}_1$ and also we expect that vector multiplet fields $X^i_v$ may acquire expectation values. The expectation values acquired by fields should lie in the quotient locus $\CM$ (see \eqref{locus}). Condition $\mu_{a,i}^{(3)}=0$ in \eqref{mm} implies that in this case fields $q_a$ can not acquire any expectation value, however the latter contradicts condition $\mu_v^{(2)}=0$ unless we also rescale $\theta_v$ as $\theta_v/{\bf s}_1$. This overall uniform re-scaling of FI parameters does not cause wall-crossing.

\subsubsection{Vertex contribution}
First of all we tend parameter ${\bf s}_2$ to infinity, this forces fields to localize on the locus $\left(\mu_{v,ij}^{(1)}\right)^{-1}(0)$. On this locus expectation values of $X^i$ commute. Using the gauge transform $U(n_v)$ in each quiver vertex $v$ we can rotate all $X^i_v$ to diagonal matrices. So we expand the fields as
\be
\begin{array}{cccc}
	X_v^i= & \hat X_v^i & + & \frac{1}{\sqrt{{\bf s}_2}}\tilde X_v^i\\
	& \uparrow & & \uparrow \\
	& {\rm diagonal}	& & {\rm off-diagonal} \\
\end{array}
\ee
For simplicity let us first consider the case when we have a quiver with a single node $v$ with $n_v=2$ so that all $X_v^i$ are two-by-two matrices. It turns out that the resulting Hamiltonian depends only on differences of diagonal values of $X_v^i$. So we choose the following parameterization:
\be
X_v^i=\left(\begin{array}{cc} 
x_i & \frac{1}{\sqrt{{\bf s}_2}} w_i\\
\frac{1}{\sqrt{{\bf s}_2}} \bar w_i & 0\\
\end{array} \right)
\ee
And for off-diagonal part of the fermion matrix we use the following noations:
\be
\lambda=\left( \begin{array}{cc}
	0 & \chi\\
	\rho & 0 \\
\end{array}\right)
\ee
A leading part of the Hamiltonian for field $w_i$ reads:
\be
\CH={\bf s}_2\left[-\sum\lm_{i}\p_{w_i}\p_{\bar w_i}+\sum\lm_{i} |\epsilon_{ijk}w_j x_k|^2-\bar\chi\bar\sigma^i x_i \chi+\bar\rho\bar\sigma^i x_i \rho\right]+O\left({\bf s}_2^\frac{1}{2}\right)
\ee
It is natural to choose new coordinates:
\be\nn
u_i=\epsilon_{ijk}w_j x_k
\ee
These coordinates are redundant: they satisfy a linear relation $w_iu_i=0$. So we introduce the fourth coordinate:
\be \nn
u_4=\sum\lm_{i}w_i x_i
\ee
Eventually we choose a new orthonormal basis: 
\be\nn
v_1=\frac{x_2u_1-x_1 u_2}{\sqrt{x_1^2+x_2^2}|\vec x|},\quad v_2= \frac{u_3}{\sqrt{x_1^2+x_2^2}},\quad v_3=\frac{u_4}{|\vec x|}
\ee

In these coordinates the first order Hamiltonian reads:
\be\label{Ham1}
\CH=-\sum\lm_{i=1}^3\p_{\bar v_i}\p_{v_i}+|\vec x|^2\sum\lm_{i=1}^2|v_i|^2-\bar\chi\bar\sigma^i x_i \chi+\bar\rho\bar\sigma^i x_i \rho
\ee
Notice a degree of freedom denoted by $v_3$ does not have a potential term. This happens because $v_3$ is a flat gauge direction. We can rewrite the bosonic part of gauge rotation generators as
\be
{\bf g}(\varphi)=\Tr\left[\varphi,X^i\right]\Pi^i=\frac{\sqrt{{\bf s}_2}}{2}\left[ 
\Tr\left( \varphi(\sigma_1+\I\sigma_2)\right)|\vec x|\p_{v_3}+\Tr\left( \varphi(\sigma_1-\I\sigma_2)\right)|\vec x|\p_{\bar v_3}
\right]+O\left({\bf s}_2^{-\frac{1}{2}}\right)
\ee
The resulting $G$-invariant wave function is independent of $v_3$, therefore we do not require it to be $L^2$-integrable, rather it should $L^2$-integrable with the measure $d^2v_1d^2 v_2$, and a direction associated to $v_3$ we assume to be a compact coordinate direction in the gauge group. To define the wave function one should define first fermion vacuum $|0\rangle$. We define it in such a way that it is annihilated by all bar-less operators:
\be\nn
\chi_\alpha|0\rangle=\rho_\alpha|0\rangle=0
\ee
Under these conditions it is not complicated to calculate the ground state of Hamiltonian \eqref{Ham1}:
\be\label{wf}
\Psi=C e^{-|\vec x|^{-1}|\epsilon_{ijk}w_j x_k|^2}\left(\bar z \bar\chi_{\dot 1}-(x^3-|\vec x|)\bar\chi_{\dot 2}\right)\left(\bar z \bar\rho_{\dot 1}-(x^3+|\vec x|)\bar\rho_{\dot 2}\right)|0\rangle
\ee
where $C$ is some normalization constant. The first order Hamiltonian did not have a kinetic term for fields $x_i$. So we may multiply this state by any wave function $\psi(\vec x)$ or take $C$ to be $\vec x$-dependent. Eventually this is some common $\vec x$-dependent multiplier of $\Psi$. We will give a prescription to distinguish $\psi$ and $C$ momentarily, now let us define the second order condition to fix their values. The composite wave function $|\psi\Psi\rangle$ belongs to a null-space of the Hamiltonian and of the supercharges. Thus we have conditions:
\be
\langle \psi' \Psi|Q_\alpha|\psi\Psi\rangle=\langle \psi' |Q_\alpha^{\rm eff}|\psi\rangle=0,\quad \langle \psi' \Psi|Q_\alpha|\psi\Psi\rangle=\langle \psi' |\bar Q_{\dot\alpha}^{\rm eff}|\psi\rangle=0
\ee
So the ground state wave function $\psi$ is annihilated by effective supercharges:
\be
Q_\alpha^{\rm eff}=\langle\Psi|Q_{\alpha}|\Psi\rangle=\I(\sigma^k\bar\lambda)_{\alpha}\left(\p_{x^k}+B_k\right) \\ 
\bar Q_{\dot \alpha}^{\rm eff}=\langle\Psi|\bar Q_{\dot\alpha}|\Psi\rangle=\I(\lambda\sigma^k)_{\alpha}\left(\p_{x^k}-B_k\right)
\ee
where for an effective Berry connection we have:
\be
B_k=\frac{\langle \Psi|\p_{x^k}|\Psi\rangle}{\langle \Psi|\Psi\rangle}
\ee
Notice that a choice of another normalization $C'$ induces a gauge transform of the Berry connection:
\be
B_k \to B_k+\p_{x^k}\log C'-\p_{x^k}\log C
\ee
So by choosing the normalization $C$ in wave function \eqref{wf} we will fix the gauge of the Berry connection. It turns out in this case the Berry connection is gauge equivalent to a trivial connection so we choose $C$ in such a way that $B_k=0$, thus we have $C=|\vec x|^{-2}\bar z^{-1}$, where $z=x_1+\I x_2$. Eventually, we define a vertex wave function corresponding to a pair of diagonal expectation values ${\vec x}_{v,i}$ and ${\vec x}_{v,j}$  for the vector multiplet in vertex $v$ as
\be\label{vertex}
\begin{split}
\mathscr{V}_{v,ij}=\frac{1}{|\vec x_{v,i}-\vec x_{v,j}|^2(\bar z_{v,i}-\bar z_{v,j})}  \exp\left[-|\vec x_{v,i}- \vec x_{v,j}|^{-1}|\vec w \times (\vec x_{v,i}- \vec x_{v,j})|^{-1})|^2\right]\times\\ 
\times \left(\left(\bar z_{v,i}- \bar z_{v,j} \right) \bar\chi_{v,ij,\dot 1}-\left((x^3_{v,i}-x^3_{v,j})-|\vec x_{v,i}-\vec x_{v,j}|\right)\bar\chi_{v,ij,\dot 2}\right)\times\\
\times \left(\left(\bar z_{v,i}- \bar z_{v,j} \right) \bar\rho_{v,ij,\dot 1}-\left((x^3_{v,i}-x^3_{v,j})+|\vec x_{v,i}-\vec x_{v,j}|\right)\bar\rho_{v,ij,\dot 2}\right)|0\rangle
\end{split}
\ee
Such wave function is a basic building block of the vertex contribution to a generic wave function.

We will need properties of this function with respect to the action of the generator $J_3+R$ to calculate PSCs \eqref{PSC} in what follows. The wave function is decomposable in a $J_3+R$-invariant part and a anti-holomorphic part:
\be
\mathscr{V}_{v,ij}=\mathscr{V}_{v,ij,(J_3+R)-{\rm inv}}\mathscr{V}_{v,ij,\overline{\rm holo}},\quad \mathscr{V}_{v,ij,\overline{\rm holo}}=(\bar z_{v,i}-\bar z_{v,j})^{-1}
\ee
Also it is important to notice that the large gauge transform in the node $v$ permutes $z_{v,i}$ and $z_{v,j}$ and simultaneously transforms fermions:
\be
\chi_{ij}\mapsto \rho_{ji},\quad \rho_{ij} \mapsto \chi_{ji}
\ee
Obviously, $\mathscr{V}_{v,ij,(J_3+R)-{\rm inv}}$ is an invariant of large gauge transformations.

\subsubsection{Arrow contribution}
In a similar way we construct an arrow contribution. Again to produce a building block wave function it is enough to consider a simplest non-trivial quiver with an arrow:
\be\nn
\begin{array}{c}
	\begin{tikzpicture}
	\draw (0,0) circle (0.2);
	\node at (0,0) {$1$};
	\draw (2,0) circle (0.2);
	\node at (2,0) {$1$};
	\draw[->] (0.2,0) -- (1.8,0);
	\end{tikzpicture}
\end{array}
\ee
Here numbers in nodes denote the dimension vector. The arrow Lagrangian depends only on the difference vector $\vec x_w-\vec x_v$, so for brevity in this subsection we will use the following notation $\vec x:=\vec x_w-\vec x_v$. And we have just a single field $q$. Since in this phase field $q$ does not acquire an expectation value we decompose it around $q=0$ critical point redefining $q\to q/\sqrt{{\bf s}_1}$. The first order supercharges read:
\be
\begin{split}
Q_{\alpha}=\I\sqrt{2{\bf s}_1}\left(
\psi_{\alpha}\pi-2\I\bar q x^i (\sigma^{i 0}\psi)_{\alpha}
\right)+O({\bf s}_1^0) \\
\bar Q_{\dot\alpha}=-\I\sqrt{2{\bf s}_1}\left(
\bar\psi_{\dot\alpha}\bar \pi+2\I(\bar\psi\bar\sigma^{0i})_{\dot\alpha}x^iq
\right)+O({\bf s}_1^0)\\
\CH={\bf s}_1\left(-\p_{\bar q}\p_q+|\vec x|^2|q|^2-\bar \psi \bar\sigma^i x^i\psi \right) +O({\bf s}_1^0)
\end{split}
\ee
The ground state wave function annihilated by these supercharges reads:
\be\label{wf2}
\Psi=Ce^{-|\vec x||q|^2}\left(\bar z \bar\psi_{\dot 1}-(x_3-|\vec x|)\bar\psi_{\dot 2} \right)|0\rangle
\ee
This state labels an element of equivariant cohomology of the quiver representation (compare to \cite{Denef}). Suppose $x_3=0$ then we could choose $Q_1$ as an equivariant Dolbeault $\p$-operator \cite{Cordes:1994fc,EqCoh}. One should just identify fermions and forms:
\be\nn
\psi_1\rightsquigarrow dq,\quad \psi_2\rightsquigarrow\frac{\p}{\p d\bar q},\quad \bar\psi_{\dot 1}\rightsquigarrow \frac{\p}{\p dq},\quad \bar\psi_{\dot 2}\rightsquigarrow d\bar q,
\ee
so that
\be
Q_1=\p_q+ \iota_{\bar \xi},
\ee
where $\xi$ is a vector field generated by equivariant rotation of $q$ with parameter $\bar z$.
To map state \eqref{wf2} to a form we should choose another fermion vacuum $|\tilde 0\rangle$, such that it is annihilated by $\iota$
$$
|\tilde 0\rangle:=\bar\psi_{\dot 1}|0\rangle
$$
In this case we have:
\be
\Psi=C e^{-|z||q|^2}(\bar z+|z|dq\wedge d\bar q)=C\;\bar z\; e^{\frac{|z|}{z}\left\{Q_1, \I q d\bar q \right\}}\sim \bar z
\ee
So this wave function is just homotopic to a Thom representative of the corresponding Euler class \cite[chapter 10]{Cordes:1994fc}.

Again we can choose a proper normalization $C$ for wave function \eqref{wf2} by fixing the Berry connection in the effective supercharges:
\be
Q_\alpha^{\rm eff}=\langle\Psi|Q_{\alpha}|\Psi\rangle=\I(\sigma^k\bar\lambda)_{\alpha}\left(\p_{x^k}+B_k\right)+\lambda_\alpha D \\ 
\bar Q_{\dot \alpha}^{\rm eff}=\langle\Psi|\bar Q_{\dot\alpha}|\Psi\rangle=\I(\lambda\sigma^k)_{\alpha}\left(\p_{x^k}-B_k\right)+\bar\lambda_{\dot\alpha} D 
\ee
Here we are able to produce a non-trivial $D$-term, since there is an expectation value
\be
\langle|q|^2\rangle=\frac{1}{|\vec x|}
\ee
And for the Berry connection we have:
\be
\vec\nabla \times \vec B=\frac{\I}{2}\nabla\frac{1}{|\vec x|}
\ee
The Berry connection correpsonds to a vector potential produced by a monopole. We choose such normalizations that in a northern(+)/southern(-) hemisphere it is given by a canonical expressions:
\be
\vec B_{\pm}=-\frac{\I}{2}\left(\pm 1-\frac{x_3}{|\vec x|} \right)\frac{1}{x_1^2+x_2^2}\left(
\begin{array}{c}
	-x_2 \\ x_1 \\ 0 \\
\end{array}
 \right)
\ee
The corresponding normalizations differ depending on whether we require the vector potentials to be regular in either northern or southern hemisphere:
\be
C_+=|\vec x|^{\frac{1}{2}}\left(|\vec x|+x_3 \right)^{-\frac{1}{2}},\quad
C_-=|\vec x|^{\frac{1}{2}}\left(|\vec x|+x_3 \right)^{-\frac{1}{2}}\left(\frac{z}{\bar z} \right)^{\frac{1}{2}}
\ee
Eventually we define a generic building block for arrow contribution:
\be\label{arrow}
\begin{split}
\mathscr{A}_{a:(v,i)\to (w,j)}=\left(\frac{|\vec x_{w,j}-\vec x_{v,i}|}{|\vec x_{w,j}-\vec x_{v,i}|+(x_{w,j}^3-x_{v,i}^3)} \right)^{\frac{1}{2}}
\left( \frac{z_{w,j}-z_{v,i}}{\bar z_{w,j}-\bar z_{v,i}} \right)^{\frac{1}{2}\Theta\left(x_{v,i}^3-x_{w,j}^3\right)}\times\\
\times e^{-|\vec x_{v,i}-\vec x_{w,j}||q_{a,ij}|^2}\left[\left(\bar z_{w,j} -\bar z_{v,i}\right)\bar\psi_{\dot 1,a, ij}-\left((x_{w,j}^3-x_{v,i}^3)-|\vec x_{w,j}-\vec x_{v,i}| \right) \bar\psi_{\dot 2,a, ij} \right]|0\rangle
\end{split}
\ee
where $\Theta$ is the Heaviside step-function.

We will need again properties of this function with respect to the action of the generator $J_3+R$ to calculate PSCs \eqref{PSC} in what follows. The wave function is decomposable in a $J_3+R$-invariant part and a anti-holomorphic part:
\be
\mathscr{A}_{a,ij}=\mathscr{A}_{a,ij,(J_3+R)-{\rm inv}}\mathscr{A}_{a,ij,\overline{\rm holo}},\quad \mathscr{A}_{a,ij,\overline{\rm holo}}=(\bar z_{w,j} -\bar z_{v,i})^{\Theta(x_{w,j}^3-x_{v,i}^3)}
\ee
In this case the anti-holomorphic part corresponds to just the Euler character mentioned before.

\subsubsection{Effective molecular description}

Summarizing the preparations performed in the previous subsections we construct an approximate wave function in the following form:
\be\label{wf3}
\Psi\left[\mathscr{Q}, \vec n,\vec \theta \right]=\psi_{\rm mol}\left(
\prod\lm_{v\in\CV(\mathscr{Q})}\prod\lm_{\substack{i,j=1\\ i<j }}^{n_v} \mathscr{V}_{v,ij}
\right)
\left(
\prod\lm_{(a:v\to w)\in \CA(\mathscr{Q})}\prod\lm_{i=1}^{n_v}\prod\lm_{j=1}^{n_w}\mathscr{A}_{a,ij}
\right)
\ee
Where the molecular part $\psi_{\rm mol}$ of the wave function is annihilated by effective supercharges corresponding to a system of mutually interacting dyons (see \cite{Denef} for comparison):
\be
\begin{split}
Q_\alpha=\sum\lm_{v\in\CV}\sum\lm_{i=1}^{n_v}\left[-\I(\sigma^0\bar\lambda_{v,i})_{\alpha}D_{v,i}-(\vec \sigma \bar\lambda_{v,i})_\alpha\left(-\I\vec\nabla_{v,i} -\vec A_{v,i} \right)\right]\\
\bar Q_{\dot\alpha}=\sum\lm_{v\in\CV}\sum\lm_{i=1}^{n_v}\left[\I(\lambda_{v,i}\sigma^0)_{\alpha}D_{v,i}-(\lambda_{v,i} \vec \sigma)_\alpha\left(-\I\vec\nabla_{v,i} -\vec A_{v,i} \right)\right]
\end{split}
\ee
Effective $D$-terms define a collection of relations corresponding to Denef equations:
\be
D_{v,i}=\theta_v+\sum\lm_{w\in\CV}\left\{\sum\lm_{a:v\to w}\sum\lm_{j=1}^{n_w}\frac{1}{|\vec x_{v,i}-\vec x_{w,j}|}-\sum\lm_{a:w\to v}\sum\lm_{j=1}^{n_w}\frac{1}{|\vec x_{v,i}-\vec x_{w,j}|}
\right\}
\ee
And the vector potential $\vec A_\alpha$ is one corresponding to the dyonic interaction and satisfying:
\be
\vec\nabla_\alpha D_\beta=\vec\nabla_\beta D_\alpha=\frac{1}{2}\left(
\vec\nabla_\alpha \times \vec A_\beta+ \vec\nabla_\beta \times \vec A_\alpha
\right)
\ee

This system describes a collection of elementary structureless dyons \cite{Denef}. To each quiver vertex $v$ one associates $n_v$ dyons of the same charge vector $\gamma_v$ taking values in an integral charge lattice $\Gamma$. Now the quiver data my be encoded in a choice of an antisymmetric lattice DSZ pairing product $\langle\cdot,\cdot\rangle$ satisfying a condition:
\be
\langle \gamma_v,\gamma_w \rangle=\#(a:v\to w)-\#(a:w\to v),
\ee
where we are counting arrows flowing between corresponding quiver nodes.
Then fields $\vec x_{v,i}$ parameterize dyon locations in $\IR^3$.

To calculate PSC \eqref{PSC} we need to insert the corresponding Wilson loop. Insertion of a Wilson loop for $J_3+R$ is analogous to turning on an $\Omega$-background like in \cite{Ohta:2014ria}, or a constant magnetic field parameterized by field strength $\eta>0$ directed along the third axis in $\IR^3$. We will mimic this insertion by modifying derivatives:\footnote{The same deformation can be mimiced by turning on $\Omega_{\eta}$-background or by modifying the height function $h_1\to h_1+\eta\Tr Z\bar Z$. This deformation affects the vertex contribution \eqref{vertex} as well, however it is of order $\eta$ so we neglect it in what follows.}
\be\nn
\p_z\to \p_z+\eta \bar z, \quad \p_{\bar z}\to \p_{\bar z}-\eta z
\ee

We follow approach of \cite{MPS} and localize the states on fixed points of the $J_3+R$-generator: to the third axis in $\IR^3$. During localization vector fields $\vec x_{v,i}$ develop vevs along the third axis we will denote as $x^*_{v,i}$. These values satisfy Denef equations:
\be\label{Denef}
\theta_v+\sum\lm_{w\in\CV}\left\{\sum\lm_{a:v\to w}\sum\lm_{j=1}^{n_w}\frac{1}{|x_{v,i}^*- x_{w,j}^*|}-\sum\lm_{a:w\to v}\sum\lm_{j=1}^{n_w}\frac{1}{|x_{v,i}^*-x_{w,j}^*|}
\right\}=0
\ee
An expansion of the supercharges around this critical point reads:
\be\label{Q_1}
\begin{split}
Q_1=\I\sum\lm_{v,i}\bar\lambda^{\dot 1}_{v,i}\left(\p_{x_{v,i}^3}-\sum\lm_{w,j}{\rm sign}(x_{v,i}^*-x_{w,j}^*)\left(\sum\lm_{a:v\to w}\frac{x_{v,i}^3-x_{w,j}^3}{|x_{v,i}^*-x_{w,j}^*|^2}- \sum\lm_{a:w\to v}\frac{x_{v,i}^3-x_{w,j}^3}{|x_{v,i}^*-x_{w,j}^*|^2} \right)  \right)+\\ + \I\sum\lm_{v,i}\bar\lambda^{\dot 2}_{v,i}\left(2\p_{z_{v,i}}+2\eta \bar z_{v,i} -\frac{1}{2}\sum\lm_{w,j}{\rm sign}(x_{v,i}^*-x_{w,j}^*)\left(\sum\lm_{a:v\to w}\frac{\bar z_{v,i}-\bar z_{w,j}}{|x_{v,i}^*-x_{w,j}^*|^2}- \sum\lm_{a:w\to v}\frac{\bar z_{v,i}-\bar z_{w,j}}{|x_{v,i}^*-x_{w,j}^*|^2} \right)  \right)
\end{split}
\ee
The last term in the second bracket mimics the contribution of the mutual electro-magnetic field. Dyon electro-magnetic field is of a point-like source, so near sphere poles it is directed along the third axis and has a magnitude $\sim 1/r^2$. This field acts as an additional correction to the constant field $\eta$ we have introduced. A problem of counting Hamiltonian eigen states in a plane with a turned on magnetic field perpendicular to this plane has a text book solution in terms of Landau levels (LL) \cite{Landau}.
When $\theta_v$ go to zero the state approaches a border of a stability region. The bound dyon molecules acquire gigantic physical sizes \cite{Galakhov:2013oja}. The mutual electro-magnetic interaction becomes negligible in comparison to external field $\eta$. Therefore finally we get a molecular wave function in \eqref{wf3} approximated by Hall states:
\be
\psi_{\rm mol}=\psi(x_{v,i}^3)\prod\lm_{v\in\CV}\prod\lm_{i=1}^{n_v}\bar z_{v,i}^{\ell_{v,i}}e^{-\eta z_{v,i}\bar z_{v,i}}|0\rangle
\ee
Here LLs are parameterized by $\ell_{v,i}\in\IZ_{\geq 0}$.

This wave function is again decomposable in a $J_3+R$-invariant part and an anti-holomorphic part given by a Hall state:
\be
\psi_{\rm mol}=\psi_{(J_3+R)-{\rm inv}}\psi_{\overline{\rm holo}},\quad \psi_{\overline{\rm holo}}=\prod\lm_{v\in\CV}\prod\lm_{i=1}^{n_v}\bar z^{\ell_{v,i}}_{v,i}
\ee
$\psi_{(J_3+R)-{\rm inv}}$ is definitely not an $L^2$-integrable function. This function should be invariant under overall translations of the center of mass of the dyon system along the third axis, so the potential has flat directions. In principle, for non-primitive dimensional vectors one expects contributions of multi-dimensional states, so the quiver representation moduli space becomes singular with branches of various dimensions. However it is invariant under $J_3+R$-symmetry and it does not contribute the index except a rather subtle fermion sign.

As it was noticed in \cite{MPS,MPS2} Denef equations \eqref{Denef} describe critical points of a height function:
\be\label{W}
W=\sum\lm_{v\in\CV}\theta_v\sum\lm_{i=1}^{n_v}x_{v,i}+\sum\lm_{(a:v\to w)\in\CA} \sum\lm_{i=1}^{n_v} \sum\lm_{j=1}^{n_w}{\rm sign}\left(x_{v,i}-x_{w,j} \right) \log|x_{v,i}-x_{w,j}|
\ee

The corresponding wave function $\psi(x_{v,i}^3)$ is given by a co-chain  in a Morse complex \cite{Witten} generated by critical points of height function $W$ we have denoted as $x^*$. The corresponding Morse co-chain we will denote as 
$$
\Lambda(x^*)=\bigwedge\lm_{\alpha}dx_{\alpha}^{\Theta(-\omega_\alpha)},
$$
where $\omega_\alpha$ are eigen values of the Hessian $\p_{x_{v,i}}\p_{x_{w,j}}W|_{x^*}$ and $x_{\alpha}$ are corresponding eigen vectors.

\subsection{Explicit wave functions}\label{wave}

As we have seen so far a generic ground state wave function can be decomposed as
\be
\Psi_{\rm mol}=\Psi_{(J_3+R)-{\rm inv}}\Psi_{\overline{\rm holo}}
\ee
The invariant part does not contribute to the counting of PSCs.
This wave function and critical points \eqref{Denef} has a natural action of the symmetric group $\prod\lm_v S_{n_v}$ acting in nodes. This action corresponds to large gauge $G$-transformations. So physical wave functions correspond to symmetrizations of found localized wave functions. When counting the BPS index one may think of it as an equivariant sum with respect to the action of this symmetric group that reduces to a sum over fixed points.\footnote{We will comment more on this in what follows.} In fixed points $\Psi_{(J_3+R)-{\rm inv}}$ part does not contribute to the index only $\Psi_{\overline{\rm holo}}$ does. Therefore for a construction of PSCs it is enough to consider only a contribution of $F$- and $(J_3+R)$-non-invariant part: the anti-holomorphic part that is given by symmetrization:
\be\label{holowf}
\begin{split}
H_{0,\overline{\rm holo}}\left(\mathscr{Q},\vec n,\vec \theta\right)=\bigoplus_{x^*;\,\ell_{v,i}\geq 0}\IC{\mathop{\rm Sym}\lm_{\prod\lm_v S_{n_v}}} \Lambda(x^*)\left( \prod\lm_{v\in\CV}\left(\prod\lm_{i=1}^{n_v}\bar z_{v,i}^{\ell_{v,i}} \prod\lm_{\substack{i,j=1\\ i<j}}^{n_v}(\bar z_{v,i}-\bar z_{v,j})^{-1}\right)\right)\times\\ \times\left( \prod\lm_{(a:v\to w)\in\CA}\prod\lm_{i=1}^{n_v}\prod\lm_{j=1}^{n_w}(\bar z_{w,j} -\bar z_{v,i})^{\Theta(x_{w,j}^*-x_{v,i}^*)}
\right)
\end{split}
\ee
Where the sum runs over solutions to Denef equations \eqref{Denef} and Landau levels, and a symmetrization goes over permutations of variables in each vertex. Then the index \eqref{PSC} reduces to a trace over only this anti-holomorphic states:
\be
\Omega(y)=(-1)^{|\CA|+|\CV|-1}y^{\sum\lm_v\frac{n_v(n_v-1)}{2}+|\CA|-1}(y-y^{-1})\times \Tr_{H_{0,\overline{\rm holo}}}\; (-1)^Fy^{-2\CJ},\quad \CJ=\sum\lm_{v,i} \bar z_{v,i}\p_{\bar z_{v,i}}
\ee
Obviously, the major counting part of this bulky expression is $\Tr_{H_{0,\overline{\rm holo}}}\; (-1)^Fy^{-2\CJ}$, and this term gives just the corresponding Poincar\'e polynomial. To match it with the PSC  we have added a multiplier $(y-y^{-1})$ corresponding to the center of mass contribution that should be canceled, and overall monomial factors so that the resulting expression corresponds to a PSC that is invariant under transformation $\Omega(y)=\Omega(y^{-1})$ for smooth quiver moduli spaces.

We will demonstrate how such index could be calculated in section \ref{MPS_form}.

Here we also should stress that $H_{0,\overline{\rm holo}}$ is not the actual Hilbert space $H_0$ of ground states. $H_{0,\overline{\rm holo}}$ is only quasi-isomorphic to $H_0$: only their indices, or Euler characteristics, coincide. To construct the actual space $H_0$ following \cite{Witten} we should construct a Morse differential 
$$
d_{\rm Morse}:\quad H_{0,\overline{\rm holo}}[f]\longrightarrow H_{0,\overline{\rm holo}}[f+1]
$$ 
saturated by steepest descend flows with respect to the height functions. The resulting $H_0$ is defined as a cohomology:
\be
H_0=H^*\left(H_{0,\overline{\rm holo}},d_{\rm Morse} \right)
\ee
In Appendix \ref{halo} we will demonstrate in a simple example of a Hall halo that the action of differential $d_{\rm Morse}$ is, in general, non-trivial.

\subsection{Manschot-Pioline-Sen index formula}\label{MPS_form}

A nice way of calculating an index of \eqref{holowf} was proposed in \cite{MPS,MPS2,Manschot:2011xc} and is based on an interplay between Fermi-Dirac, Bose-Einstein and Boltzman statistics. We could notice that the symmetrization appeared in \eqref{holowf} corresponds to a remnant of integration over the gauge group, and as we mentioned the index we are computing corresponds to an equivariant integral. Therefore the index reduces to a sum over fixed points. The fixed points of the symmetric group are labeled by partitions $\{(l)^{k_l} \}$. For example, an equivariant character for symmetric group is given by:
\be\label{eq_sym}
\chi_{\rm sym}(n)=\sum\lm_{\substack{\{k_l \} \\ \sum\lm_l l k_l=n}} \left(\prod\lm_l\frac{1}{k_l! l^{k_l}} \right)\chi\left(\{(l)^{k_l} \}\right)
\ee
where $\chi$ is a character without imposing any symmetrization.
As the simplest demonstration of this concept we could introduce an equivariant character on polynomials of $n$ variables by weighting each monomial by fugacity $q^{\rm power}$. Then the character of symmetric polynomials in $n$ variables reads
$$
\chi_{\rm sym}(n)=\prod\lm_{j=1}^n\frac{1}{1-q^j}
$$
Fixed points of the symmetric group are monomials with coincident powers. For example,  fixed points of $S_3$ are:
$$
\{3\}:\; z_1^k z_2^k z_3^k,\quad \{2,1 \}:\; z_1^{k_1} z_2^{k_1} z_3^{k_2},\quad \{1,1,1 \}:\; z_1^{k_1} z_2^{k_2} z_3^{k_3}
$$
Corresponding character are:
$$
\quad \chi\left(\{(l)^{k_l} \}\right)=\prod\lm_l\frac{1}{(1-q^l)^{k_l}}
$$
And for these characters relation \eqref{eq_sym} holds.

So using \eqref{eq_sym} we relate characters for various statistics.

\begin{figure}[h!]
	\begin{center}
		\begin{tikzpicture}
		\node(A) at (0,0) {$
			\begin{array}{c}
			\begin{tikzpicture}
			\draw (0,0) circle (0.2) (1.5,0) circle (0.2);
			\node at (0,0) {$2$};
			\node at (1.5,0) {$2$};
			\draw[->] (0.2,0) -- (1.3,0);
			\node[above] at (0.75,0) {$\varkappa$};
			\node[above] at (0,0.2) {$\theta_1$};
			\node[above] at (1.5,0.2) {$\theta_2$};
			\end{tikzpicture}
			\end{array}
			$};
		\node(B) at (5,0) {$
			\begin{array}{c}
			\begin{tikzpicture}
			\draw (0,0.4) circle (0.2) (1.5,0.4) circle (0.2) (0,-0.4) circle (0.2) (1.5,-0.4) circle (0.2);
			\node at (0,0.4) {$1$};
			\node at (1.5,0.4) {$1$};
			\node at (0,-0.4) {$1$};
			\node at (1.5,-0.4) {$1$};
			\draw[->] (0.2,0.4) -- (1.3,0.4);
			\draw[->] (0.2,-0.4) -- (1.3,-0.4);
			\draw[->] (0.17,0.3) -- (1.33,-0.3);
			\draw[->] (0.17,-0.3) -- (1.33,0.3);
			\node[above] at (0.75,0.4) {$\varkappa$};
			\node[above] at (0,0.6) {$\theta_1$};
			\node[above] at (1.5,0.6) {$\theta_2$};
			\end{tikzpicture}
			\end{array}
			$};
		\node(C) at (10,0) {$
			\begin{array}{c}
			\begin{tikzpicture}
			\draw (0,0) circle (0.2) (1.5,0.4) circle (0.2) (1.5,-0.4) circle (0.2);
			\node at (0,0) {$1$};
			\node at (1.5,0.4) {$1$};
			\node at (1.5,-0.4) {$1$};
			\draw[->] (0.17,-0.1) -- (1.31,-0.35);
			\draw[->] (0.17,0.1) -- (1.31,0.35);
			\node[above] at (0.75,0.3) {$2\varkappa$};
			\node[below] at (0.75,-0.3) {$2\varkappa$};
			\node[above] at (0,0.2) {$2\theta_1$};
			\node[right] at (1.7,0.4) {$\theta_2$};
			\node[right] at (1.7,-0.4) {$\theta_2$};
			\end{tikzpicture}
			\end{array}
			$};
		\path (A) edge[->] node[above]{split} (B) (B) edge[->] node[above]{$\begin{array}{c} {\rm fixed\; point}\\ (\{2\} , \; \{1 ,1 \}) \end{array}$} (C);
		\end{tikzpicture}
	\end{center}
	\caption{Split quiver}\label{SQ}
\end{figure}
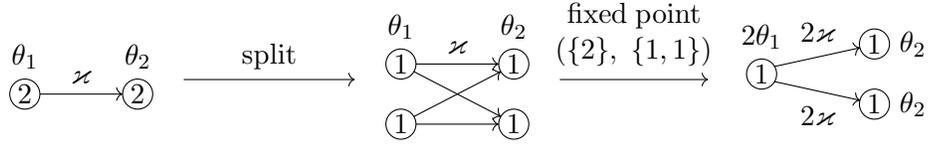

The Boltzmanian index corresponding to the Hilbert space \eqref{holowf} without symmetrization procedure is easy to compute. The corresponding index and Denef equations \eqref{Denef} correspond to an Abelian split quiver. We can ``Abelianize" a node $v$ with corresponding dimension $n_v$ by splitting it in $n_v$ nodes of dimension 1 with preserving all arrows and FI parameters (see Figure \ref{SQ}). A split quiver corresponds to a quiver with all nodes Abelianized.

The Boltzmanian index for quiver $\mathscr{Q}$ is given just by a sum over LLs and a weight from the arrow contribution for a given fixed point of Denef equations:
\be
\Omega_{\rm Boltzman}(y,\mathscr{Q} )=(-1)^{|\CA|+|\CV|-1}y^{\sum\lm_v\frac{n_v(n_v-1)}{2}+|\CA|-1}\left(\frac{1}{y-y^{-1}}\right)^{\sum\lm_v n_v-1}\sum\lm_{x^*}(-1)^{f(x^*)}y^{-2\sum\lm_a\#(a:v\to w)\Theta(x_{w,j}^*-x_{v,i}^*)}
\ee
where $f$ is a corresponding fermion number given by a half of difference of numbers of positive and negative eigen values of the Hessian of $W$.

The fixed points of the symmetric group correspond to gluing nodes of the quiver with summing arrows and corresponding FI parameters according to partitions (see Figure \ref{SQ}). As well when gluing quiver vertices we should identify corresponding LLs:
$$
\bar z_1^{\ell_1}\bar z_2^{\ell_2}\to \bar z_1^{\ell}\bar z_2^{\ell}
$$
Therefore the index contribution of a single LL is modified. We should add corresponding $y$-factors. The result is the Manschot-Pioline-Sen formula for quiver index without loops and with prime dimension vectors:
\be\label{MPS_formula}
\Omega_{\rm MPS}(y,\mathscr{Q})=\sum\lm_{\substack{\{\vec k_l \} \\ \sum\lm_l l \vec k_l=\vec n}} \left(\prod\lm_i\prod\lm_l\frac{1}{k_l^{(i)}! l^{k_l^{(i)}}}\left(\frac{y-y^{-1}}{y^l-y^{-l}} \right)^{k_l^{(i)}} \right)\Omega_{\rm Boltzman}\left(y,\mathscr{Q}\left(\{(l)^{\vec k_l} \}\right) \right)
\ee
where the sum runs over partition vectors corresponding to partitioning each component of the dimension vector $\vec n$. This formula works for quivers without loops and primitive dimension vectors. For a more generic situation the counting should be more careful \cite{MPS2}.

\section{Cohomological Hall algebra}\label{CoHA}
\subsection{Framed BPS states}\label{f_states}

\begin{figure}[h!]
	\begin{center}
		\begin{tikzpicture}
		\draw (-3,0) circle (0.1) (-1,0) circle (0.1) (1,0) circle (0.1) (3,0) circle (0.1);
		\draw[->] (-2.9,0) -- (-1.1,0);
		\draw[->] (2.9,0) -- (1.1,0);
		\draw[dashed] (-0.9,0) -- (0.9,0);
		\draw[->] (0,-2) edge node[below left] {$k$} (-2.93,-0.07);
		\draw[->] (0,-2) edge node[below right] {$k$} (2.93,-0.07);
		\draw[->] (0,-2) edge node[left] {$k$} (-0.97,-0.09);
		\draw[->] (0,-2) edge node[right] {$k$} (0.97,-0.09);
		\draw[fill=red] (-0.1,-1.9) -- (0.1,-1.9) -- (0.1,-2.1) -- (-0.1,-2.1) -- cycle; 
		\node[right] at (0.1,-2.1) {framing node};
		\end{tikzpicture}
	\end{center}
	\caption{Framed quiver} \label{frm_quiv}
\end{figure}
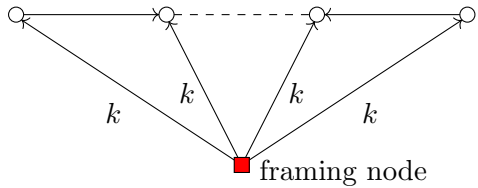

A calculation of generic PSCs even using MPS formula \eqref{MPS_formula} is a hard combinatorial problem. To count BPS states and corresponding indices a much more efficient way is to use various wall-crossing relations. A natural way to approach such relations is to use a probe in a class $\CS$ theory -- a line defect \cite{Gaiotto:2010be,Gukov:2006jk}. Line defects define  proper supersymmetric monodromy conditions around 1d manifolds  for fields in a theory. A nice family of such defects is given by generalized Wilson-'tHooft lines. In a low energy effective picture such defects may be represented by heavy dyons, so in the quiver language they are translated to additional framing nodes added to a quiver. A framing node corresponds to a real mass term contribution to the Lagrangian \eqref{Lag} (see, for example, \cite{Chung:2016pgt}). In other words, we should consider an ordinary quiver vertex, give to the vector field $X^3$ a scalar vev, say, $x_c$ and put all the oher fields to zero. We could choose different framings, for our further purposes we choose a framing node connected to all the other quiver vertices by $k$ outgoing arrows (see Figure \ref{frm_quiv}). Notice that in this way we do not add oriented loops to the quiver.

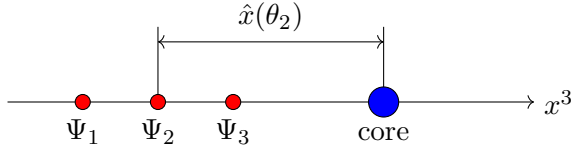
\begin{figure}[h!]
	\begin{center}
		\begin{tikzpicture}
		\draw[->] (0,0) -- (7,0);
		\draw (2,0) -- (2,1) (5,0) -- (5,1);
		\draw[<->] (2,0.8) -- (5,0.8);
		\node[above] at (3.5,0.8) {$\hat x(\theta_2)$};
		\node[right] at (7,0) {$x^3$};
		\draw[fill=red] (1,0) circle (0.1) (2,0) circle (0.1) (3,0) circle (0.1);
		\draw[fill=blue] (5,0) circle (0.2);
		\node[below] at (5,-0.2) {core};
		\node[below] at (1,-0.1) {$\Psi_1$};
		\node[below] at (2,-0.1) {$\Psi_2$};
		\node[below] at (3,-0.1) {$\Psi_3$};
		\end{tikzpicture}
	\end{center}
	\caption{Framed BPS molecule}\label{f_BPS}
\end{figure}

In the language of BPS molecules we get a heavy core dyon located at $x_c$ on the third axis of $\IR^3$. We will go first to a regime when the very BPS molecule is localized in a neighborhood of $\hat x$. Then changing coordinates by shifting $x_{v,i}\to x_{v,i}+\hat x$ we observe that the system of Denef equations \eqref{Denef} is decomposed in a system of two approximate equation sets. The first one corresponds to a sum over all equations, then all the mutual dyon interaction terms cancel each other and we end up with contributions of FI terms and framing terms. The framing terms we approximate by a contribution when all the dyons are put in one point $\hat x$ 
\be
\sum\lm_v n_v\theta_v-\frac{k\sum\lm_v n_v}{|x_c-\hat x|}=0
\ee
This equation has two solutions when the state is localized to the right or the left side of the framing core. For a reason that will be clear further we concentrate only on the left critical point (see Figure \ref{f_BPS}).

And the corresponding molecule localization point is given by the follwoing expression:
\be\label{BPS_loc}
\hat x=x_c-\frac{k\sum\lm_{v} n_v}{\sum\lm_{v} n_v \theta_v}
\ee
Substituting this expression back to the Denef equations we will get another approximate system of unframed equations with redefined stability parameters having now a zero average:
\be
\tilde \theta_v=\theta_v-\frac{\sum\lm_{v}n_v\theta_v}{\sum\lm_{v}n_v}
\ee
Therefore the wave function of the light dyons is represented by an ordinary wave function for an unframed quiver we have calculated before with effective stability parameters $\tilde \theta_v$ and localized near $\hat x$.

The FI parameters entering the effective 1d $\CN=4$ quiver SQM are related to phases of complex central charges of the elementary dyons in the effective theory \cite{Gaiotto:2010be}:
\be\label{sp}
\theta_v=2|\CZ_c|^{-1}{\rm Im}\left(\bar \CZ_c \CZ_v \right)
\ee
where $\CZ_c$ is a large central charge of the core. The denominator in the BPS molecule location formula \eqref{BPS_loc} reads
$$
{\rm Im}(\bar \CZ_c \CZ_{\mu})
$$
where 
$$
\CZ_{\mu}=\sum\lm_v n_v \CZ_v
$$
is a total central charge of a molecule. 

Suppose all the central charges lie in the upper half-plane. Then if $\CZ_c$ lies to the right of $\CZ_{\mu}$, $\theta_{\mu}>0$ there is a critical point and the BPS molecule is bound to the core. As $\CZ_c$ approaches $\CZ_{\mu}$ the binding radius $r\sim \left({\rm Im}(\bar \CZ_c \CZ_{\mu})\right) ^{-1}$ becomes greater and greater until the framed bound state dissolves completely:
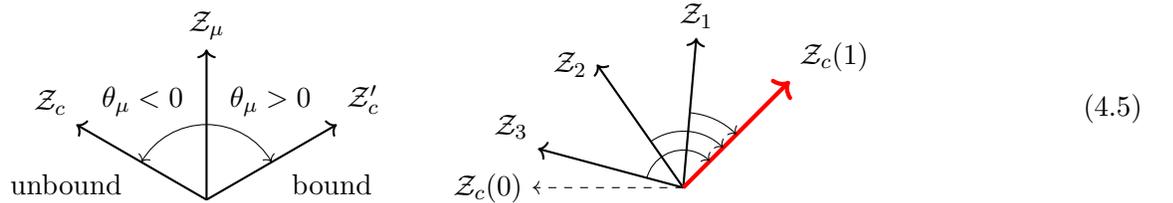
\be
\begin{array}{c}
	\begin{tikzpicture}
	\draw[thick,->] (0,0) -- (0,2);
	\node[above] at (0,2) {$\CZ_\mu$};
	\begin{scope}[rotate=60]
	\draw[thick,->] (0,0) -- (0,2);
	\node[above left] at (0,2) {$\CZ_c$};
	\end{scope}
	\begin{scope}[rotate=-60]
	\draw[thick,->] (0,0) -- (0,2);
	\node[above right] at (0,2) {$\CZ_c'$};
	\end{scope}
	\draw[->] ([shift=(90:1)]0,0) arc (90:150:1);
	\draw[->] ([shift=(90:1)]0,0) arc (90:30:1);
	\node[left] at (-1,0.2) {unbound};
	\node[right] at (1,0.2) {bound};
	\begin{scope}[rotate=-10]
	\node[above right] at (0,1) {$\theta_{\mu}>0$};
	\end{scope}
	\begin{scope}[rotate=10]
	\node[above left] at (0,1) {$
		\theta_{\mu}<0$};
	\end{scope}
	\end{tikzpicture}
\end{array}\quad
\begin{array}{c}
	\begin{tikzpicture}
	\begin{scope}[xscale=-1]
	\begin{scope}[rotate=15]
	\draw[thick,->] (0,0) -- (2,0);
	\node[above left] at (2,0) {$\CZ_3$};
	\begin{scope}[rotate=40]
		\draw[thick,->] (0,0) -- (2,0);
		\node[left] at (2,0) {$\CZ_2$};
	\end{scope}
	\begin{scope}[rotate=80]
		\draw[thick,->] (0,0) -- (2,0);
		\node[above] at (2,0) {$\CZ_1$};
	\end{scope}
	\begin{scope}[rotate=120]
	\draw[red,ultra thick,->] (0,0) -- (2,0);
	\node[above right] at (2,0) {$\CZ_c(1)$};
	\end{scope}
	\draw[->] ([shift=(0:0.5)]0,0) arc (0:120:0.5);
	\draw[->] ([shift=(40:0.75)]0,0) arc (40:120:0.75);
	\draw[->] ([shift=(80:1)]0,0) arc (80:120:1);
	\end{scope}
	\draw[dashed,->] (0,0) -- (2,0);
 	\node[left] at (2,0) {$\CZ_c(0)$};
 	\end{scope}
	\end{tikzpicture}
\end{array}\label{CZ_c}
\ee

Now assume there is an adiabatic change of $\CZ_c(t)$ with some adiabatic parameter $t\in[0,1]$, so that it crosses some molecular $\CZ_{\mu}$ in a clockwise order. The corresponding molecular states may appear from an asymptotic infinity and scatter on the core, the mutual interaction between scattering molecules is of order $\sim k^{-1}$, so assuming $k\gg 1$ we can neglect it. The order of appearance of these molecules coincides with the order of corresponding $\theta_{\mu}$ or counterclockwise order of central charges (see Figure \ref{f_BPS}):
\be
\hat x_1<\hat x_2\; \Leftrightarrow \; \theta_1<\theta_2
\ee

Let us construct a scattering asymptotic state. When there is just one BPS molecule in a state $\Psi$ the scattering BPS state is also $\Psi$. Indeed in the wave function \eqref{holowf} we may expect a contribution from arrows connecting the quiver vertices with the framing node. However since we have chosen to scatter BPS molecules from the left this contribution is cancelled by Heaviside $\Theta$-functions in \eqref{holowf}. Suppose now we are scattering two consequently appearing BPS molecule states $\Psi_1$ and $\Psi_2$. The resulting wave function we will call a product of states $\Psi_1\cdot \Psi_2$. It is just given by a multiplication by corresponding factors in \eqref{holowf} and taking symmetrization:
\be\label{product}
\Psi_1\cdot\Psi_2\left(\{\bar z_{v,i}\}\cup\{\bar z_{v,i}'\} \right)=\sum\lm_{\rm shuffles}\Psi_1\left(\{\bar z_{v,i}\} \right)\Psi_2\left(\{\bar z_{v,i}'\} \right)\frac{\prod\lm_{v,w\in\CV}\prod\lm_{i=1}^{n_1^v}\prod\lm_{j=1}^{n_2^w} 
	(\bar z_{v,i}-\bar z_{w,j}')^{\#(a:v\to w)}}{\prod\lm_{v\in\CV}\prod\lm_{i=1}^{n_1^v} \prod\lm_{j=1}^{n_2^v}(\bar z_{v,i}-\bar z_{v,j}' )}
\ee
The shuffles correspond to symmetrization in \eqref{holowf} over $\prod_{v}S_{n_v}$ for $\vec n=\vec n_1+\vec n_2$ applied to a joint set of variables $\{\bar z_{v,i}\}\cup\{\bar z_{v,i}'\}$.
Let us here notice that this product is obviously non-commutative, the order of the wave functions is defined by localization positions of corresponding BPS molecules in $\IR^3$ and, therefore, by BPS molecule stability parameter averages $\theta_{1,2}=\vec n_{1,2}\cdot\vec\theta$.

We would like to argue that multiplication \eqref{product} is cohomological: maps cohomologies $H_0$ to cohomologies $H_0$. As we mentioned in section \ref{wave} an element on the right hand side of \eqref{product} is an element of $H_{0,\overline{\rm holo}}$, and there is generic Morse differential that can be represented by the following expansion:
\be
d_{\rm Morse}=d_1\otimes 1+1\otimes d_2+d_{\rm mix}
\ee
where $d_{1,2}$ are differentials acting only on states $\Psi_{1,2}$ correspondingly, and $d_{\rm mix}$ is a differential mixing two BPS molecules. $d_{\rm mix}$ is saturated by instanton processes when some of elemantary dyons get splitted from one BPS molecule cluster and join another one. According to \cite{Witten} such instanton amplitude defining matrix elements of $d_{\rm mix}$ is given by exponentiated difference of height function \eqref{W} values in initial and final states, so it is dominated by the following contribution:
\be\nn
e^{\Delta W}\sim e^{-\alpha|\hat x_1-\hat x_2|}
\ee
where $\alpha>0$ is some combination of $\theta_v$ and $\hat x_{1,2}$ are localization coordinates of BPS clusters 1 and 2. Since we consider these states in asymptotic limit when $|\hat x_1-\hat x_2|\to\infty$ the instanton amplitude iz zero and we conclude:
$$
d_{\rm mix}=0
$$
So the differential commutes with multiplication \eqref{product}.

\subsection{Scattering algebra}
In \cite{Moore} Harvey and Moore proposed a natural construction of an algebra for BPS states:
\be
m_{\rm HM}:\quad H_{\rm BPS}\otimes H_{\rm BPS}\to H_{\rm BPS}
\ee
promoting S-matrices to structure constants of this algebra. Let us try to formulate this statement more precisely. Suppose we scatter two BPS states and get a new possibly multiparticle state $\Psi_3$. Central charges during this process are summed, so $\CZ_3=\CZ_1+\CZ_2$. If there is a bound state of particles 1 and 2 the scattering amplitude as a complex function of Mandelstam $s$-variable has a distinguished pole:
\be\label{scatt}
\CA(12\to 3)=\frac{\langle\Psi_3|\Psi_1\cdot \Psi_2\rangle}{s-|\CZ_1+\CZ_2|^2}
\ee
The residue at this pole defines a product in the algebra $\Psi_1\cdot\Psi_2$. Unfortunately, we will be unable to apply this idea directly in our situation. A reason for that is we have approximated states that are exact BPS states, so we can not deviate from the physical locus where $s$ is real and satisfies inequality
$$
s= \left(|\CZ_1|+|\CZ_2|\right)^2 >|\CZ_1+\CZ_2|^2
$$
unless we sit exactly at the marginal stability (MS) locus where $|\CZ_1|+|\CZ_2|=|\CZ_1+\CZ_2|$. Let us consider a small deviation from the MS locus parameterized by $t$, so that one hits the MS wall at $t=0$:
$$
|\CZ_1(t)|+|\CZ_2(t)|>|\CZ_1(t)+\CZ_2(t)| \;\;\mbox{for}\;\; t>0,\; \;\mbox{and}\;\; |\CZ_1(0)|+|\CZ_2(0)|=|\CZ_1(0)+\CZ_2(0)|
$$
We get a family of $\Psi_3(t)$ varying adiabatically with $t$.
The fact that value $t=0$ corresponds to an MS wall is reflected in the behavior of $\Psi_3(t)$: as $t$ approaches the value $t=0$ the state 3 becomes unstable and its wave function gets factorized in two (or more) single particle asymptotic wave functions:
\be\nn
\lim\lm_{t\to 0}\Psi_3(t)= \mathscr{S}_{12}^3\; \Psi_{1}\otimes \Psi_2
\ee
So we have defined an adibatic S-matrix limit:
\be\nn
\mathscr{S}_{12}^3=\lim\lm_{t\to 0}\langle\Psi_3(t)|\Psi_1 \Psi_2\rangle
\ee
In the case when $\Psi_3(t)$ corresponds to more than two asymptotic single particle states we assume this matrix element is zero.
Using this expression as a pole residue in \eqref{scatt} we define scattering algebra of BPS wave functions as:
\be
\Psi_1\cdot \Psi_2=\mathscr{S}_{12}^3\Psi_1\Psi_2
\ee

We can describe types and numbers of the dyons belonging to the first and the second scattered states by charge vectors $\vec n_1$ and $\vec n_2$. Multiplication will map a product of these wave functions to a wave function corresponding to a charge vector $\vec n_1+\vec n_2$. It is natural to grade the anti-holomorphic Hilbert spaces by charge vectors. Elements of these spaces -- Hall wave functions -- together with the multiplication form an algebra we could call a Scattering algebra for Hall states (SAHS). As we argued in the previous subsection this algebra is cohomological. The differential factors through the multiplication structure, so we can use \eqref{product} as a product rule for $H_0$:
\be
m_{\rm SAHS}: \quad H_{(0,\;\vec n_1)}\otimes H_{(0,\;\vec n_2)}\to H_{(0,\;\vec n_1+\vec n_2)}
\ee
An explicit formula for $m_{\rm SAHS}$ we have constructed in section \ref{f_states}, it is given by relation \eqref{product}.

A subtlety of this construction we should mention here is that $\mathscr{S}_{12}^3$ for \eqref{product} is not just a number rather a rational function of $\bar z$'s. This is not so surprising in comparison to QFT text book scattering process examples since state $\Psi_3$ depends also on degrees of freedom of a field agent whose negative field energy generates a bound for states $\Psi_1$ and $\Psi_2$. Usually, we consider an asymptotic state of $\Psi_1$ and $\Psi_2$ with a mutual interaction turned off. So the sate $|\Psi_1 \Psi_2\rangle$ is approximated in the first order in coupling constant as a product of three ingredients: states $\Psi_1$, $\Psi_2$ and a binding agent put in the ground state. In our situation we are unable to put the binding agent --  mutual electro-magnetic field in the effective description --  in the ground state since it carries a non-trivial angular momentum for a dyonic pair given by DSZ pairing.

As we mentioned in Section \ref{s31} it is natural to identify the ground states with equivariant cohomologies of quiver representations. Associating to supercharge $Q_1$ action of the differential we find that the Weil algebraic element is parameterized by fields $\bar Z_v$. And the equivariant cohomologies $H^*$ correspond \cite{KS} to polynomials in variables $\bar z_{v,i}$ parameterizing the equivatriant torus. So it is natural to identify spaces of anti-holomorphic wave functions and $H^*$. $H^*$ can also be graded by charge vectors $\vec n$. Elements of $H^*$ form a cohomological Hall algebra (CoHA) with a multiplication introduced by Kontsvich and Soibelman \cite{KS}:
\be
m_{\rm CoHA}:\quad H^*_{\vec n_1}\otimes H^*_{\vec n_2}\to H^*_{\vec n_1+\vec n_2}
\ee
Using a straightforward comparison we observe:
\be
m_{\rm SAHS}=m_{\rm CoHA}
\ee

Therefore we conclude that the scattering algebra of Hall states and cohomological Hall algebra are equivalent. Finally we could draw some conclusions from this equivalence. SAHS and CoHA are associative algebras and together correspond to a special case of Feigin-Odesskii shuffle algebra \cite{Feigin}. 

\subsection{Wall-crossing formul{\ae}}

Wall-crossing formulae is a simple set of relations giving a description of how PSCs vary as we vary stability parameters. The basic idea of constructing these relations is an invariance of the framed BPS states under variation of the stability parameters. Wall-crossing phenomena are represented by BPS molecule decays or recombinations, however as we have showed in the section \ref{f_states} a framed BPS molecule remains bound while central charges $\CZ_i$ remain inside the cone formed by $\CZ_c(0)$ and $\CZ_c(1)$ (see \eqref{CZ_c}). Even if we reshuffle the very central charges $\CZ_i$. Shuffles of $\CZ_i$ lead to shuffles of an ordered set of corresponding stability parameters \eqref{sp} allowing one to move between various stability regions. Comparing framed BPS wave functions one can recover what wave functions of stable unframed BPS states have contributed in various stability regions.

So suppose we have scattered $N$ unframed BPS states defined by a set of dimension vectors $\{\vec n^{(\alpha)} \}_{\alpha=1}^N$. The the framed BPS wave function is constructed as
\be
\Psi_{\rm framed}=\prod\lm_{\{\vec n^{(\alpha)} \}}^{\curvearrowright} \Psi_{\vec n^{(\alpha)}}
\ee
Where as a product formula we apply associative product \eqref{product}, and the states are ordered according to the following rule:
\be
\alpha<\beta,\quad \theta^{(\alpha)}<\theta^{(\beta)}
\ee
where
$$
\theta^{(\alpha)}=\frac{\sum\lm_v n_v^{(\alpha)}\theta_v}{\sum\lm_v n_v^{(\alpha)}}
$$
Here we should notice that dimensional vectors $\vec n$ and $k\vec n$, $k\geq 2$ have the same net stability parameters $\theta^{(\alpha)}$. Therefore they are localized at the same position in $\IR^3$ or, equivalently, at the same ray of central charges. Therefore it is natural to combine Hilbert spaces of various dyon numbers in Fock spaces of dyons when we sum up over all multipliers of a primitive dimensional vector:
\be
\CF_{\vec n}=\bigoplus\lm_{k=1}^{\infty}H_{k\vec n}
\ee

Then for Fock spaces we have the following wall-crossing formul{\ae}. The product
$$
\bigotimes\lm_{\vec n}^{\curvearrowright} \CF_{\vec n}
$$
remains invariant under reshuffles of the stability parameters.

A dramatic and still simple example can be given in terms of a Kronecker quiver with one arrow. In \cite{KS} it was shown that:
\be
\CF_{(1,0)}\otimes \CF_{(0,1)}\cong \CF_{(0,1)}\otimes \CF_{(1,1)}\otimes \CF_{(1,0)}
\ee 
On the left hand side we have an ordering of parameters $\theta_1<\theta_2$, and there is no boundstate, therefore only Fock spaces corresponding to elementary dyons contribute. On the right hand side the ordering of the stability parameters is changed to $\theta_1>\theta_2$, and a Fock space of a boundstate with charge vector $(1,1)$ appears. We will review this example in more details in Appendix \ref{Wall-crossing}.

It turns out to be quite useful to use indices for whole Fock spaces. In addition to the usual definition \eqref{PSC} we could add a fugacity for the total dyonic charge $\CQ$, as a result we will get characters of Fock spaces:
\be
\Phi(u,y)=\Tr_{\CF}(-1)^F y^{-2\CJ}u^{\CQ}
\ee

To take into account non-commutativity of the product \eqref{product} let us notice that for two states with charge vectors $\vec n$ and $\vec n'$ we have
\be
\CJ(\Psi_{\vec n}\cdot \Psi_{\vec n'})-\CJ(\Psi_{\vec n'}\cdot \Psi_{\vec n})=\langle\vec n,\vec n'\rangle
\ee
where the right hand side is given by anti-symmetric DSZ pairing on the charge lattice:
\be
\langle\vec n,\vec n'\rangle:=\sum\lm_{v,w}n_v n_w'\left(\#(a:v\to w)- \#(a:w\to v) \right)
\ee
Or equivalently,
$$
y^{-2\CJ}(\Psi_{\vec n}\cdot \Psi_{\vec n'})=y^{-2\langle\vec n,\vec n'\rangle}y^{-2\CJ}(\Psi_{\vec n'}\cdot \Psi_{\vec n})
$$

So we assume that fugacities inherit non-commutativity property of the algebra, so when we permute them in the product they produce corresponding $y$-factor from permutation of states:
\be
u_{\vec n}u_{\vec n'}=y^{-2\langle\vec n,\vec n'\rangle}u_{\vec n'}u_{\vec n}= y^{-\langle\vec n,\vec n'\rangle}u_{\vec n+\vec n'}
\ee

Let us calculate such indices for some simple examples. We will start with a single node quiver. The corresponding Fock space reads:
$$
\CF_{0}=\bigoplus\lm_{n\geq 1}\bigoplus\lm_{0\leq \ell_1<\ldots\ell_n}\IC{\rm Sym}\frac{\bar z_1^{\ell_1}\cdots \bar z_n^{\ell_n}}{\prod\lm_{i<j}(\bar z_i-\bar z_j)}
$$
We will have contributions to the index from the numerator and the denominator of these wave functions. The contribution from the denominator depends only on the number of particles $n$ and can be canceled by a reparameterization of the $u$-fugacity and leads to an overall $y$-factor of the index. In principle, only relative degrees of various LL contributions matter. We choose a normalization of indices (overall $y$-factor) as in $\cite{KS}$ so that to a state labeled by LLs $0\leq\ell_1<\ldots<\ell_n$ we associate degree $\sum\lm_i\ell_i-n$. The corresponding index of the Fock space reads:
\be
\Phi_o(u,y)=\sum\lm_{k=0}^{\infty}\frac{y^{-k^2}}{\prod\lm_{j=1}^k(1-y^{-2j})}u^k=(-y^{-1}u;y^{-2})_{\infty}
\ee
where we used a conventional notation for Pochhammer symbols (or quantum dilogarithm functions):
$$
(x;q)_{\infty}:=\prod\lm_{j=0}^{\infty}(1-x q^j)
$$
If we have a generic BPS molecule with an index given by a finite Laurent polynomial in $y$:
\be
\Omega(y)=\sum\lm_{s}a_s y^s
\ee
we assume that it is given by contributions of states $\psi_{s,p}$ so that we have $p=1,\ldots,a_s$ states of spin projection $s$. When BPS molecule is bound to a defect core we should add a contribution of a center of mass degree of freedom, so the corresponding framed BPS wave functions would read:
$$
\psi_{s,p} \bar z^{\ell}
$$
These states may be combined in multi-molecule contributions:
$$
\psi_{s,p} \bar z^{\ell_1}\cdot \psi_{s,p} \bar z^{\ell_2}\cdot\ldots
$$
If states are analogous to hypermultiplets they are anti-commuting. A contribution of a Fock space of multiple $\psi_{s,p}$-components is just $\Phi_o(y^{s}u,y)$. On the other hand, if states are commuting the corresponding contribution is $\Phi_o(-y^{s}u,y)^{-1}$. The no-exotics theorem \cite{Chuang:2013wt} predicts that there are no exotics states, so PSCs correspond to a sum of characters of spin $SU(2)$ irreps with either integer or half-integer spins. The former ones correspond to anti-commuting states, the latter ones correspond to commuting states since in the definition we have extracted the contribution of a hypermultiplet corresponding to the center of mass degree of freedom.

So, eventually, a contribution of a Fock space of a BPS molecule with index $\Omega(y)$ reads \cite{Gaiotto:2010be}:
\be
\Phi^{\Omega}(u,y)=\prod\lm_{s}\Phi_o\left({\rm sign}(a_s)y^{s}u,y\right)^{a_s}
\ee
Applying the generating function to an ordered product of Fock spaces we derive that an ordered product over primitive dimension vectors:
\be\label{wall-crossing}
\prod\lm_{\vec n}^{\curvearrowright}\Phi^{\Omega_{\vec n}(y)}(u_{\vec n},y)={\rm const}
\ee
remains invariant under  reshuffles of stability parameters. Statement \eqref{wall-crossing} establishes Kontsevich-Soibelman wall-crossing relations \cite{Kontsevich:2008fj,Gaiotto:2010be} for motivic Donaldson-Thomas invariants.

\appendix
\section{Examples}\label{examples}

\subsection{Hall halo}\label{halo}

As a simple example let us consider a Hall halo molecule.

\begin{figure}[h!]
	\begin{center}
		\begin{tikzpicture}
		\draw (0,0) circle (1.5);
		\draw[fill=red] (0,1.5) circle (0.08) (0,-1.5) circle (0.08);
		\begin{scope}[shift={(0,1.5)},yscale=0.5]
		\draw[->] ([shift=(45:0.3)]0,0) arc (45:415:0.3);  
		\end{scope}
		\begin{scope}[shift={(0,-1.5)},yscale=0.5]
		\draw[->] ([shift=(45:0.3)]0,0) arc (45:415:0.3); 
		\end{scope}
		\begin{scope}[yscale=0.3]
		\draw ([shift=(180:1.5)]0,0) arc (180:360:1.5);
		\draw[dashed] ([shift=(0:1.5)]0,0) arc (0:180:1.5);
		\end{scope}
		\draw[fill=blue] (0,0) circle (0.15);
		\node[above] at (0,1.6) {$\mathscr{N}$};
		\node[below] at (0,-1.6) {$\mathscr{S}$};
		\node[below] at (0,1.4) {$\psi_N$};
		\node[above] at (0,-1.4) {$\psi_S$};
		\end{tikzpicture}
	\end{center}
	\caption{Hall halo BPS molecule}\label{HH}
\end{figure}
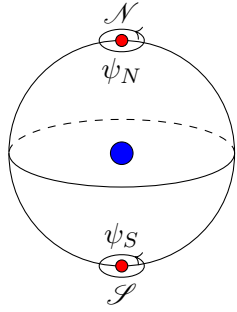

The Hall halo is a standard molecular configuration when there is a single heavy magnetic center with charge $\varkappa$ and $n<\varkappa$ electrically charged light electrons floating around and forming a ``halo" (see Figure \ref{HH}).
The corresponding quiver is given by the following diagram:
\be\nn
\begin{array}{c}
	\begin{tikzpicture}
	\draw (0,0) circle (0.2);
	\node at (0,0) {$1$};
	\draw (2,0) circle (0.2);
	\node at (2,0) {$n$};
	\draw[->] (0.2,0) -- (1.8,0);
	\node[above] at (1,0) {$\varkappa$};
	\end{tikzpicture}
\end{array}
\ee

The system of Denef equations \eqref{Denef} has a solution only if $-\theta_2=n\theta_1=\theta>0$. In this case all the light dyons are confined in the poles of a sphere of radius $\theta^{-1}$.

Let us first consider the case when $n=1$. The Hilbert space consists of two subspaces where the light particle is localized in either the north($+$) or the south pole ($-$). The anti-holomophic wave function in the south pole is related to the function in the north pole as
\be
\psi_S(\ell_1;\ell_2)=[1] \sum\lm_{i=0}^{\varkappa}\left(\begin{array}{c} \varkappa\\ i \end{array} \right) \psi_N(\ell_1+i;\ell_2+\varkappa-i)
\ee
By $[1]$ we imply a fermion number shift. The height function \eqref{W} has a single flat modulus corresponding to the center of mass degree of freedom. Therefore the north pole Hessian of $W$ has a single positive eigen value and a zero one, the south pole one has a negative and zero eigenvalues. The fermion number shift appears in this way. It is natural to split the center of mass degree of freedom, so we impose $\ell_2=0$. The wave function reduces to 
\be\label{NS}
\psi_S(\ell_1;0)=[1]  \psi_N(\ell_1+\varkappa;0)
\ee
In the index all the north pole wave functions with $\ell_1\geq\varkappa$ will be canceled with corresponding south pole  wave functions. We conclude there is the only contribution to the index from $n$ electrons confined in the north pole with LLs $\ell<\varkappa$. The light dyons localized in the north pole have anti-commuting wave functions due to denominators in \eqref{holowf}. Therefore we conclude the anti-holomorphic Hilbert space is isomorphic to:
\be
H_{0,\overline{\rm holo}}\cong\bigoplus\lm_{0\leq \ell_1<\ell_2<\ldots<\ell_n<\varkappa}\IC\; \psi_{N,{\rm anti-symm}}(\ell_1,\ell_2,\ldots,\ell_n;0)
\ee
The corresponding index reads:
\be\label{ind}
\Omega(y)=(-1)^{\varkappa-1}\frac{[\varkappa]_y!}{[n]_y![\varkappa-n]_y!}
\ee

In this example of a Hall halo molecule a difference between anti-holomorphic Hilbert space $H_{0,\overline{\rm holo}}$ and actual Hilbert space of ground states $H_0$ becomes transparent. Index \eqref{ind} we have calculated is just a $y$-deformed binomial coefficient ``$\varkappa$ choose $n$". It is well-known that a surface of finite area with magnetic flux $\varkappa$ through it has a finite number of Landau levels of order $\varkappa$ \cite{Landau}, in our particular case the number of LLs is exactly given by $\varkappa$. Wave functions corresponding to these Landau levels were explicitly calculated in \cite{Denef}. So the problem of counting ground states in this model reduces to a problem to distribute $n$ equivalent particles with fermion statistics between $\varkappa$ Landau levels. Corresponding Hilbert space of ground states $H_0$ has exactly dimension $\varkappa$ choose $n$. However anti-holomorphic Hilbert space $H_{0,\overline{\rm holo}}$ is infinite dimensional. To find an explicit subspace $H_0\subset H_{0,\overline{\rm holo}}$ one should calculate a Morse differential $d_{\rm Morse}$ corresponding to non-perturbative instanton corrections \cite{Witten}. Then the actual ground space Hilbert space is a Morse cohomology:
$$
H_0\cong H^*\left(H_{0,\overline{\rm holo}}, d_{\rm Morse} \right)
$$
It is not complicated to calculate an explicit action of this differential following \cite{Witten,Hori:2003ic}, in particular $d_{\rm Morse}=Q_1$, and $Q_1$ has a non-trivial matrix elements on $H_{0,\overline{\rm holo}}$ saturated by instantons -- trajectories fixed by the action of the supercharge. In this case SUSY transformations \eqref{SUSY}
of fermions read:
\be\nn
\delta \lambda=\I \xi \left(\theta+\frac{1}{|\vec x|}\right)+2\sigma^{0i}\xi \dot x_i
\ee
After applying the Wick rotation we easily see that the instanton equation reads:
$$
\dot x_3=-\left(\theta+\frac{1}{|\vec x|}\right),\quad \dot z=0
$$
The corresponding instanton trajectory is just a straight vertical line connecting northern and southern hemispheres. A particle travels along this line for an infinite time, therefore the instanton has a single modulus corresponding to time translations. As it is explained in \cite{Hori:2003ic} this instanton modulus corresponds by the supersymmetry to a zero fermion mode in the instanton background canceling the fermion insertion of operator $Q_1$ in the path integral.  Since the coordinate $z$ in the horizontal plane remains unchanged along the instanton trajectory, the corresponding transition amplitude is proportional to the delta function $\delta(z_{\rm south}-z_{\rm north})$. We conclude that 
$$
d_{\rm Morse}\psi_N =\psi_N[1],\quad d_{\rm Morse}\psi_S =0,
$$
so in cohomology exactly cancellation \eqref{NS} occurs.

\subsection{$1$-Kronecker quiver wall-crossing from CoHA}\label{Wall-crossing}
Here let us review in greater details an example of wall crossing in the CoHA presented in \cite{KS}. We consider a 1-Kronecker quiver:
\be\nn
\begin{array}{c}
	\begin{tikzpicture}
	\draw (0,0) circle (0.2);
	\draw (2,0) circle (0.2);
	\draw[->] (0.2,0) -- (1.8,0);
	\end{tikzpicture}
\end{array}
\ee
The algebra is generated in general by two infinite setes of variables $\xi_i$ and $\eta_i$, $i\geq 0$ corresponding to Hall one particles monomials $\bar z^i$ for each quiver vertex.

This algebra reads:
\be
\xi_i\xi_j+\xi_j\xi_i=0,\quad \eta_i\eta_j+\eta_j\eta_i=0,\quad \eta_i\xi_j=\xi_{j+1}\eta_i-\xi_{j} \eta_{i+1}
\ee

As we have seen in Appendix \ref{halo} a state of charge $(1,1)$ is a singlet, and it may be approximated by a single wave function $\xi_0\eta_0$. In this wave function we have suppressed contribution of the molecule center of mass. By restoring it we expect wave functions of type $\nu_i=\xi_i\eta_0$ or $\tilde \nu_i=\xi_0\eta_i$. Both $\nu_i$ and $\tilde\nu_i$ form antisymmetric algebras:
\be
\nu_i\nu_j+\nu_j\nu_i=0,\quad \tilde\nu_i\tilde\nu_j+\tilde\nu_j\tilde\nu_i=0
\ee
And the polynomial rings $\IZ[\nu_i]$ and $\IZ[\tilde\nu_i]$ are isomorphic. Each one represents a single copy of the $(1,1)$ molecule Fock space.
Therefore boundstates $(1,1)$ behave themselves mechanically as ordinary elementary dyon hypermultiplets as we expected.

Wall-crossing relation for an ordered product of Fock spaces is translated in the following relation between polynomial rings:
\be
\IZ[\xi_i]\cdot\IZ[\eta_i]\cong \IZ[\eta_i]\cdot\IZ[\xi_i\eta_0]\cdot\IZ[\xi_i]
\ee
Here let us write down explicitly some generators of lower degree from the both sides of this relation and establish an isomorphism between them.

Generators of charge $(1,1)$ and degree 0 are:
\be
\begin{split}
{\rm LHS}: \xi_0\cdot \eta_0=\xi_0\eta_0\\
{\rm RHS}: 1\cdot\xi_0 \eta_0\cdot 1=\xi_0\eta_0\\
\end{split}
\ee

Generators of charge $(1,1)$ and degree 1 are:
\be
\begin{split}
	{\rm LHS}: \xi_1\cdot \eta_0=\xi_1\eta_0,\quad \xi_0\cdot \eta_1=\xi_0\eta_1\\
	{\rm RHS}: 1\cdot\xi_1 \eta_0\cdot 1=\xi_1\eta_0,\quad \eta_0\cdot 1\cdot \xi_0=\xi_1\eta_0-\xi_0\eta_1\\
\end{split}
\ee

Generators of charge $(1,1)$ and degree 2 are:
\be
\begin{split}
	{\rm LHS}: \xi_2\cdot \eta_0=\xi_2\eta_0,\quad \xi_1\cdot \eta_1=\xi_1\eta_1,\quad \xi_0\cdot \eta_2=\xi_0\eta_2\\
	{\rm RHS}: 1\cdot\xi_2 \eta_0\cdot 1=\xi_2\eta_0,\quad \eta_1\cdot 1\cdot \xi_0=\xi_1\eta_1-\xi_0\eta_2, \quad \eta_0\cdot 1\cdot \xi_1=\xi_2\eta_0-\xi_1\eta_1\\
\end{split}
\ee

Generators of charge $(2,1)$ and degree 3 are:
\be
\begin{split}
	{\rm LHS}: \xi_0\xi_3\cdot \eta_0=\xi_0\xi_3 \eta_0,\quad \xi_1\xi_2\cdot \eta_0=\xi_1\xi_2\eta_0,\quad \xi_0\xi_2\cdot \eta_1=\xi_0\xi_2\eta_1,\quad \xi_0\xi_1\cdot \eta_2=\xi_0\xi_1 \eta_2\\
	{\rm RHS}: \eta_0\cdot 1\cdot \xi_0\xi_1=\xi_1\xi_2\eta_0-\xi_0\xi_2\eta_1+\xi_0\xi_1\eta_2,\quad 1\cdot \xi_0\eta_0\cdot \xi_2=\xi_0\xi_3\eta_0-\xi_0\xi_2\eta_1,\\
	1\cdot \xi_1\eta_0\cdot \xi_1=\xi_1\xi_2\eta_0,\quad 1\cdot\xi_2\eta_0\cdot\xi_0=\xi_0\xi_2\eta_1-\xi_1\xi_2\eta_0
\end{split}
\ee
In all these cases there is an obvious isomorphism between LHS and RHS bases.

\subsection{$\varkappa$-Kronecker quiver: non-primitive dimension vector $(2,2)$}
In the case of non-primitive dimensional vector ${\rm gcd}(n_1,n_2,\ldots)>1$ the quiver representation moduli space is a non-smooth stack \cite{Reineke3,MPS2}. This obstacle can be observed on the level of Denef equations \eqref{Denef}: various types of solutions to the same system of equations have flat moduli spaces of different dimension.

To resolve this situation we again involve framing as a regularization procedure. Denoting coordinates of particles corresponding to the first node as $x_i$ and to the second node as $y_i$ we derive the following framed Denef equations:
\be\label{A_12}
\begin{split}
	\theta_1-\sum\lm_{j=1}^2\frac{\varkappa}{|x_i-y_j|}-\frac{k}{|x_i|}=0\\
	\theta_2+\sum\lm_{j=1}^2\frac{\varkappa}{|y_i-x_j|}-\frac{k}{|y_i|}=0
\end{split}
\ee 

It is easy to solve them numerically. From solutions we need only ordering of particle locations in the space and eigenspaces of the corresponding Hessian of $W$. One would expect that re-scaling $k$ and $\varkappa$ will not affect a mutual disposition of elementary dyons, so in solving these equations numerically we take $\varkappa=k=1$ and $\theta_1=1.2$, $\theta_2=-1.0$.  The result is summarized in the following table:
\be
\begin{array}{lll}
	\begin{array}{l}
		{\rm a)}\; (x,x,y,y),\; q^0\\
		\hline
		\begin{array}{l|l}
			+ & x_1+x_2-y_1-y_2\\
			+ & y_1-y_2 \\
			+ & x_1-x_2\\
			+ & x_1+x_2+y_1+y_2\\
		\end{array}\\
	\hline
	\end{array}&
	\begin{array}{l}
		{\rm c)}\; (x,y,y,x),\; q^{2\varkappa}\\
		\hline
		\begin{array}{l|l}
			+ & -\alpha x_1+x_2+y_1+y_2\\
			- &  x_1-\alpha x_2+y_1+y_2 \\
			+ & y_1-y_2\\
			+ & x_1+x_2+y_1+y_2\\
		\end{array}\\
		\hline
	\end{array}&
\begin{array}{l}
	{\rm e)}\; (y,x,y,x),\; q^{3\varkappa}\\
	\hline
	\begin{array}{l|l}
		- & \ldots\\
		- &  \ldots \\
		+ & \ldots\\
		+ & \ldots\\
	\end{array}\\
	\hline
\end{array}
\\
	\begin{array}{l}
		{\rm b)}\; (y,y,x,x),\; q^{4\varkappa}\\
		\hline
		\begin{array}{l|l}
			- & x_1+x_2-y_1-y_2\\
			- & y_1-y_2 \\
			- & x_1-x_2\\
			+ & x_1+x_2+y_1+y_2\\
		\end{array}\\
	\hline
	\end{array}&
\begin{array}{l}
	{\rm d)}\; (y,x,x,y),\; q^{2\varkappa}\\
	\hline
	\begin{array}{l|l}
		+ & -\alpha y_1+y_2+x_1+x_2\\
		- &  y_1-\alpha y_2+x_1+x_2 \\
		+ & x_1-x_2\\
		+ & x_1+x_2+y_1+y_2\\
	\end{array}\\
	\hline
\end{array}&
\begin{array}{l}
	{\rm f)}\; (x,y,x,y),\; q^{\varkappa}\\
	\hline
	\begin{array}{c}
		\\
		\mbox{NO SOLUTION}\\
		\\
		\\
	\end{array}\\
	\hline
\end{array}
\end{array}\label{table}
\ee
For each configuration we have derived signs of all Hessian eigenvalues and corresponding eigenvectors. To each negative eigenvector $\sum\lm_i c_i x_i$ one associates a 1-form $\sum\lm_i c_i dx_i$ where elementary forms $dx_i$ correspond to fermion wave functions. For example, for configuration (b) we have the following wave functions (here just for brevity we denoted corresponding ${\bar z}_i$-variables as $x_i$):
$$
\IC\mathop{\rm Sym}\lm_{\{x_1,x_2\}, \{y_1,y_2\}}\frac{x_1^{\ell_1}x_2^{\ell_2}y_1^{\ell_3}y_2^{\ell_4} \prod\lm_{i,j=1}^2(x_i-y_j)^{\varkappa}}{(x_1-x_2)(y_1-y_2)}(dx_1+dx_2-dy_1-dy_2)\wedge(dx_1-dx_2)\wedge( dy_1-dy_2)
$$
Obviously, this space is just isomorphic to a tensor product of two bases of symmetric polynomials of two variables.

One should notice that molecule configuration (f) in \eqref{table} is special. There is no converging solution to \eqref{A_12} supporting this molecule configuration. Therefore it does not contribute to the index.

Summing up contributions of a) -- e) molecular states we derive the following Poincar\'e polynomial:
\be
P_{\rm frmd}=\left(\frac{q}{(1-q)(1-q^2)}\right)^2-q^{4\varkappa} \left(\frac{1}{(1-q)(1-q^2)}\right)^2-2q^{2\varkappa}\frac{q}{(1-q)^3(1-q^2)}+\frac{q^{3\varkappa}}{(1-q)^4}
\ee
To derive the unframed BPS index for $(2,2)$-state we should subtract a contribution of a doubled $(1,1)$ molecules. Corresponding Poincar\'e polynomial reads:
$$
P_{(1,1)}=\frac{1-q^{\varkappa}}{1-q}=\sum\lm_{k}a_kq^k
$$
And we should take into account that $(1,1)$ state has a parity $(-1)^{\varkappa-1}$. So the contribution of the doubled $(1,1)$-molecule is given by coefficient in the front of $u^2$ in the Taylor expansion of corresponding dilogarithms:
$$
P'=q^{\varkappa}\left.\prod\lm_k((-1)^{\varkappa-1}q^k u;q)_{\infty}^{(-1)^{\varkappa-1} a_k}\right|_{u^2}=\left\{\begin{array}{ll}
\frac{(1-q^{\varkappa})(2q-(1+q^2)q^{\varkappa})}{(1-q)^2(1-q^2)^2}, & \varkappa\;{\rm odd}\\
\frac{(1-q^{\varkappa})((1+q^2)-2 q^{\varkappa+1})}{(1-q)^2(1-q^2)^2},
& \varkappa\;{\rm even}
\end{array} \right.
$$
We subtract a contribution of a doubled $(1,1)$ molecule:
$$
P_{(2,2)}=P_{\rm frmd}-q^{\varkappa}P'
$$
Finally, to derive the corresponding index we should substitute $q=y^{-2}$ and strip off an overall $y$-monomial factor. The result reads:

For odd $\varkappa$:
\be
\Omega_{(2,2)}(y)=\frac{(y^{\varkappa-1}-y^{-(\varkappa-1)})^2(y^{2\varkappa}-y^{-2\varkappa})}{(y-y^{-1})(y^2-y^{-2})^2}
\ee

For even $\varkappa$:
\be
\Omega_{(2,2)}(y)=\frac{(y^{\varkappa-2}-y^{-(\varkappa-2)})(y^\varkappa-y^{-\varkappa})(y^{2\varkappa}-y^{-2\varkappa})}{(y-y^{-1})(y^2-y^{-2})^2}
\ee

These PSC expressions coincide with ones produced from a functional equation for PSCs proposed in \cite{GLM}.


\begin{thebibliography}{90}
	
\bibitem{Aganagic:2010qr} 
M.~Aganagic and K.~Schaeffer,
``Wall Crossing, Quivers and Crystals,''
JHEP {\bf 1210}, 153 (2012)
[arXiv:1006.2113 [hep-th]].

\bibitem{Alim:2011kw} 
M.~Alim, S.~Cecotti, C.~Cordova, S.~Espahbodi, A.~Rastogi and C.~Vafa,
``$\mathcal{N} = 2$ quantum field theories and their BPS quivers,''
Adv.\ Theor.\ Math.\ Phys.\  {\bf 18}, no. 1, 27 (2014)
[arXiv:1112.3984 [hep-th]].
	
\bibitem{Awata:2017cnz} 
H.~Awata, H.~Kanno, A.~Mironov, A.~Morozov, A.~Morozov, Y.~Ohkubo and Y.~Zenkevich,
``Generalized Knizhnik-Zamolodchikov equation for Ding-Iohara-Miki algebra,''
Phys.\ Rev.\ D {\bf 96}, no. 2, 026021 (2017)
[arXiv:1703.06084 [hep-th]].
	
	
\bibitem{Beem:2017ooy} 
C.~Beem and L.~Rastelli,
``Vertex operator algebras, Higgs branches, and modular differential equations,''
JHEP {\bf 1808}, 114 (2018)
[arXiv:1707.07679 [hep-th]].


\bibitem{Brennan:2018yuj} 
T.~D.~Brennan, A.~Dey and G.~W.~Moore,
JHEP {\bf 1809}, 014 (2018)
doi:10.1007/JHEP09(2018)014
[arXiv:1801.01986 [hep-th]].

\bibitem{Brennan:2018ura} 
T.~D.~Brennan, G.~W.~Moore and A.~B.~Royston,
JHEP {\bf 1809}, 038 (2018)
doi:10.1007/JHEP09(2018)038
[arXiv:1805.08783 [hep-th]].


\bibitem{Bullimore:2016hdc} 
M.~Bullimore, T.~Dimofte, D.~Gaiotto, J.~Hilburn and H.~C.~Kim,
``Vortices and Vermas,''
arXiv:1609.04406 [hep-th].


\bibitem{Cecotti:2012se} 
S.~Cecotti,
``The quiver approach to the BPS spectrum of a 4d N=2 gauge theory,''
Proc.\ Symp.\ Pure Math.\  {\bf 90}, 3 (2015)
[arXiv:1212.3431 [hep-th]].

\bibitem{Chuang:2013wt} 
W.~Y.~Chuang, D.~E.~Diaconescu, J.~Manschot, G.~W.~Moore and Y.~Soibelman,
``Geometric engineering of (framed) BPS states,''
Adv.\ Theor.\ Math.\ Phys.\  {\bf 18}, no. 5, 1063 (2014)
[arXiv:1301.3065 [hep-th]].

\bibitem{Chung:2016pgt} 
H.~J.~Chung and T.~Okazaki,
``(2,2) and (0,4) supersymmetric boundary conditions in 3d $\mathcal{N}$ = 4 theories and type IIB branes,''
Phys.\ Rev.\ D {\bf 96}, no. 8, 086005 (2017)
[arXiv:1608.05363 [hep-th]].

	
\bibitem{Cirafici:2008sn} 
M.~Cirafici, A.~Sinkovics and R.~J.~Szabo,
``Cohomological gauge theory, quiver matrix models and Donaldson-Thomas theory,''
Nucl.\ Phys.\ B {\bf 809}, 452 (2009)
[arXiv:0803.4188 [hep-th]].



\bibitem{Cordes:1994fc} 
S.~Cordes, G.~W.~Moore and S.~Ramgoolam,
``Lectures on 2-d Yang-Mills theory, equivariant cohomology and topological field theories,''
Nucl.\ Phys.\ Proc.\ Suppl.\  {\bf 41}, 184 (1995)
[hep-th/9411210].

\bibitem{Cordova:2014oxa} 
C.~Cordova and S.~H.~Shao,
``An Index Formula for Supersymmetric Quantum Mechanics,''
arXiv:1406.7853 [hep-th].

\bibitem{Denef} 
F.~Denef,
``Quantum quivers and Hall / hole halos,''
JHEP {\bf 0210}, 023 (2002)
[hep-th/0206072].

\bibitem{Denef:2007vg} 
F.~Denef and G.~W.~Moore,
``Split states, entropy enigmas, holes and halos,''
JHEP {\bf 1111}, 129 (2011)
[hep-th/0702146].


\bibitem{Douglas:1996sw} 
M.~R.~Douglas and G.~W.~Moore,
``D-branes, quivers, and ALE instantons,''
hep-th/9603167.

\bibitem{Feigin} B. L. Feigin, A. V. Odesskii, ``Vector bundles on elliptic curve and Sklyanin algebras," 	arXiv:q-alg/9509021

\bibitem{Gabella:2017hpz} 
M.~Gabella, P.~Longhi, C.~Y.~Park and M.~Yamazaki,
``BPS Graphs: From Spectral Networks to BPS Quivers,''
JHEP {\bf 1707}, 032 (2017)
[arXiv:1704.04204 [hep-th]].


\bibitem{Gaiotto:2009we} 
D.~Gaiotto,
``N=2 dualities,''
JHEP {\bf 1208}, 034 (2012)
[arXiv:0904.2715 [hep-th]].


\bibitem{Gaiotto:2009hg} 
D.~Gaiotto, G.~W.~Moore and A.~Neitzke,
``Wall-crossing, Hitchin Systems, and the WKB Approximation,''
arXiv:0907.3987 [hep-th].


\bibitem{Gaiotto:2010be} 
D.~Gaiotto, G.~W.~Moore and A.~Neitzke,
``Framed BPS States,''
Adv.\ Theor.\ Math.\ Phys.\  {\bf 17}, no. 2, 241 (2013)
[arXiv:1006.0146 [hep-th]].




\bibitem{Galakhov:2013oja} 
D.~Galakhov, P.~Longhi, T.~Mainiero, G.~W.~Moore and A.~Neitzke,
``Wild Wall Crossing and BPS Giants,''
JHEP {\bf 1311}, 046 (2013)
[arXiv:1305.5454 [hep-th]].



\bibitem{GLM} 
D.~Galakhov, P.~Longhi and G.~W.~Moore,
``Spectral Networks with Spin,''
Commun.\ Math.\ Phys.\  {\bf 340}, no. 1, 171 (2015)
[arXiv:1408.0207 [hep-th]].

\bibitem{Gukov:2006jk} 
S.~Gukov and E.~Witten,
``Gauge Theory, Ramification, And The Geometric Langlands Program,''
hep-th/0612073.


\bibitem{Harvey:1995fq} 
J.~A.~Harvey and G.~W.~Moore,
``Algebras, BPS states, and strings,''
Nucl.\ Phys.\ B {\bf 463}, 315 (1996)
[hep-th/9510182].

\bibitem{Moore} 
J.~A.~Harvey and G.~W.~Moore,
``On the algebras of BPS states,''
Commun.\ Math.\ Phys.\  {\bf 197}, 489 (1998)
[hep-th/9609017].


\bibitem{Hori:2003ic} 
K.~Hori, S.~Katz, A.~Klemm, R.~Pandharipande, R.~Thomas, C.~Vafa, R.~Vakil and E.~Zaslow,
\emph{Mirror symmetry}, Vol. 1. American Mathematical Soc., 2003.

\bibitem{Kimura:2016dys} 
T.~Kimura and V.~Pestun,
``Quiver elliptic W-algebras,''
Lett.\ Math.\ Phys.\  {\bf 108}, no. 6, 1383 (2018)
[arXiv:1608.04651 [hep-th]].


\bibitem{Kontsevich:2008fj} 
M.~Kontsevich and Y.~Soibelman,
``Stability structures, motivic Donaldson-Thomas invariants and cluster transformations,''
arXiv:0811.2435 [math.AG].


\bibitem{KS} 
M.~Kontsevich and Y.~Soibelman,
``Cohomological Hall algebra, exponential Hodge structures and motivic Donaldson-Thomas invariants,''
Commun.\ Num.\ Theor.\ Phys.\  {\bf 5}, 231 (2011)
[arXiv:1006.2706 [math.AG]].


\bibitem{Landau} L.D.~Landau and E.M.~Lifshitz, \emph{Quantum mechanics}, vol. 3. Course of theoretical physics,  1977.

\bibitem{Laughlin} R.B.~Laughlin,
``Quantized Hall conductivity in two-dimensions,''
Phys.\ Rev.\ B {\bf 23}, 5632 (1981).

\bibitem{EqCoh} M.~Libine, ``Lecture Notes on Equivariant Cohomology", 	arXiv:0709.3615 [math.SG]

\bibitem{Manschot:2011xc} 
J.~Manschot, B.~Pioline and A.~Sen,
``A Fixed point formula for the index of multi-centered N=2 black holes,''
JHEP {\bf 1105}, 057 (2011)
[arXiv:1103.1887 [hep-th]].




\bibitem{MPS} 
J.~Manschot, B.~Pioline and A.~Sen,
``From Black Holes to Quivers,''
JHEP {\bf 1211}, 023 (2012)
[arXiv:1207.2230 [hep-th]].




\bibitem{MPS2} 
J.~Manschot, B.~Pioline and A.~Sen,
``On the Coulomb and Higgs branch formulae for multi-centered black holes and quiver invariants,''
JHEP {\bf 1305}, 166 (2013)
[arXiv:1302.5498 [hep-th]].


\bibitem{Melnikov:2018zfn} 
D.~Melnikov, A.~Mironov, S.~Mironov, A.~Morozov and A.~Morozov,
``From Topological to Quantum Entanglement,''
arXiv:1809.04574 [hep-th].

\bibitem{Moore:1997ar} 
G.~W.~Moore,
``String duality, automorphic forms, and generalized Kac-Moody algebras,''
Nucl.\ Phys.\ Proc.\ Suppl.\  {\bf 67}, 56 (1998)
[hep-th/9710198].

\bibitem{Moore:2017byz} 
G.~W.~Moore,
``A Comment On Berry Connections,''
arXiv:1706.01149 [hep-th].


\bibitem{Moore:2015szp} 
G.W.~Moore, A.B.~Royston and D.~Van den Bleeken,
``Semiclassical framed BPS states,''
JHEP {\bf 1607}, 071 (2016)
[arXiv:1512.08924 [hep-th]].


\bibitem{Nekrasov:2012xe} 
N.~Nekrasov and V.~Pestun,
``Seiberg-Witten geometry of four dimensional N=2 quiver gauge theories,''
arXiv:1211.2240 [hep-th].



\bibitem{Ohta:2014ria} 
K.~Ohta and Y.~Sasai,
``Exact Results in Quiver Quantum Mechanics and BPS Bound State Counting,''
JHEP {\bf 1411}, 123 (2014)
[arXiv:1408.0582 [hep-th]].


\bibitem{Ohta:2015fpe} 
K.~Ohta and Y.~Sasai,
``Coulomb Branch Localization in Quiver Quantum Mechanics,''
JHEP {\bf 1602}, 106 (2016)
[arXiv:1512.00594 [hep-th]].


\bibitem{Pestun:2016zxk} 
V.~Pestun {\it et al.},
``Localization techniques in quantum field theories,''
J.\ Phys.\ A {\bf 50}, no. 44, 440301 (2017)
[arXiv:1608.02952 [hep-th]].


\bibitem{Pioline:2011gf} 
B.~Pioline,
``Four ways across the wall,''
J.\ Phys.\ Conf.\ Ser.\  {\bf 346}, 012017 (2012)
[arXiv:1103.0261 [hep-th]].

\bibitem{Rapcak:2018nsl} 
M.~Rapcak, Y.~Soibelman, Y.~Yang and G.~Zhao,
``Cohomological Hall algebras, vertex algebras and instantons,''
arXiv:1810.10402 [math.QA].


\bibitem{Reineke1}M.~Reineke, ``Moduli of representations of quivers," arXiv:0802.2147 [math-ph].


\bibitem{Reineke2}M.~Reineke, ``Cohomology of quiver moduli, functional equations, and integrality of
Donaldson–Thomas type invariants," Compositio Mathematica {\bf 147} no. 03, (2011) 943–964



\bibitem{Reineke3}M.~Reineke, ``The Harder-Narasimhan system in quantum groups and cohomology of quiver
moduli," Invent. Math. {\bf 152} (2003), no. 2, 349-368 [arXiv:math/0204059 [math.QA]].


\bibitem{Seiberg:1994rs} 
N.~Seiberg and E.~Witten,
``Electric - magnetic duality, monopole condensation, and confinement in N=2 supersymmetric Yang-Mills theory,''
Nucl.\ Phys.\ B {\bf 426}, 19 (1994)
[hep-th/9407087].


\bibitem{Shifman1} O.~Schiffmann, ``Lectures on Hall algebras", arXiv:math/0611617 [math.RT]


\bibitem{Shifman2} O.~Schiffmann, ``Lectures on canonical and crystal bases of Hall algebras", 	arXiv:0910.4460 [math.QA]

\bibitem{SO1} O.~Schiffmann, E.~Vasserot, ``On cohomological Hall algebras of quivers : Yangians", 	arXiv:1705.07491 [math.RT]

\bibitem{SO2} O.~Schiffmann, E.~Vasserot, ``On cohomological Hall algebras of quivers : generators", 	arXiv:1705.07488 [math.RT]

\bibitem{Strominger:1996sh} 
A.~Strominger and C.~Vafa,
``Microscopic origin of the Bekenstein-Hawking entropy,''
Phys.\ Lett.\ B {\bf 379}, 99 (1996)
[hep-th/9601029].


\bibitem{WessBagger} J.~Wess, and J.~Bagger, \emph{Supersymmetry and supergravity}, Princeton university press, 1992.



\bibitem{Witten} E.~Witten,
``Supersymmetry and Morse theory,''
J.\ Diff.\ Geom.\  {\bf 17}, no. 4, 661 (1982).

\bibitem{Witten:1997sc} 
E.~Witten,
``Solutions of four-dimensional field theories via M-theory,''
Nucl.\ Phys.\ B {\bf 500}, 3 (1997)
[hep-th/9703166].


\end{thebibliography}
\end{document}